\documentstyle[sprocl,epsfig]{article}
\newcommand{\beq}{\begin{equation}}
\newcommand{\ve}{\langle \Phi\rangle}
\newcommand{\diff}{\partial}
\newcommand{\eeq}{\end{equation}}
\newcommand{\beqn}{\begin{eqnarray}}
\newcommand{\eeqn}{\end{eqnarray}}
\newcommand{\bea}[1]{\beq\begin{array}{#1}}
\newcommand{\eea}{\end{array}\eeq}

\newcommand{\ma}{monopole-antimonopole }
\newcommand{\be}{\begin{equation}}
\newcommand{\ee}{\end{equation}}
\newcommand{\La}{\Lambda_{QCD}}

\newcommand{\aQ}{\alpha_s{(Q^2)}}
\newcommand{\aM}{\alpha_s(M^2)}
\newcommand{\as}{\alpha_s}
\begin{document}

~\vspace{-5mm}
\begin{flushright}   {\large ITEP-TH-36/99}  \end{flushright}
\vspace{15mm}

\centerline{\bf{PHYSICS OF THE POWER CORRECTIONS IN QCD.
\footnote{Lectures given by V.I. Zakharov at the Winter School of
physics of ITEP, Snegiri (Moscow Region), February 1999}}}

\vspace{10mm}
\centerline{{\bf F.V. Gubarev, M.I. Polikarpov}}
\vspace{3mm}
\centerline{Institute of Theoretical and Experimental Physics,}
\centerline{
B. Cheremushkinskaya, 25, 
117259 Moscow}
\centerline{E-mail:
Fedor.Gubarev@itep.ru, polykarp@vxitep.itep.ru} 
\vspace{5mm} 
\centerline{{\bf V.I. Zakharov}}
\vspace{3mm}
\centerline{Max-Planck Institut f\"ur Physik,}
\centerline{F\"ohringer Ring 6, 80805 M\"unchen, Germany}
\centerline{ E-mail: xxz@mppmu.mpg.de} 
\vspace{10mm}

\abstracts{We review the physics of the power corrections
to the parton model.
In the first part, we consider the power corrections 
which characterize 
the infrared sensitivity of Feynman graphs when the contribution
of short distances dominates.
The second part is devoted to the hypothetical power corrections associated
with nonperturbative effects at small distances.}

\section{INTRODUCTION.}

\section*{ General remarks on the power corrections.}

We consider QCD and processes determined by physics at short distances.
Which means that there is a generic
large mass scale, $Q\gg \La$ where $\La$ is the
position of
the Landau pole in the coupling:
\be
\aQ~\approx~{1\over b_0 \; \ln{Q^2/\La^2}}\label{coupling}.
\ee
For example, $Q$ may stand for the total energy
in the $e^+e^-$ annihilation into hadrons or the 4-momentum of
the virtual photon in deep inelastic scattering (DIS).

Then one can use the perturbation theory and predictions for 
a physical observable $O$ are given by a perturbative series
\footnote{In fact, the perturbative corrections may modify
the parton model by a powers of $\aQ$ as an overall factor as well.
These are so called anomalous dimensions, best known from the example
of moments from structure functions in DIS. For simplicity
we consider the case of zero anomalous dimensions.}:
\be
<O>~=<O>_{parton~model}\left(1+\sum_{n=1}^{\infty}a_n\aQ^n~\right).
\ee
Now, we reserve for powers of $\La/Q$ as well:
\be
<O>~=<O>_{parton~model}\left(1+\sum_{n=1}^{\infty}a_n\aQ^n~
+\sum_{n=k}^{\infty}b_n(\La/Q)^n\right)\label{two}\ee
These are the power corrections.

First of all, the power corrections appear to be a pure 
nonperturbative effect. Indeed, on one hand we have 
\be
\left({\La\over Q}\right)^k~=~\exp(-k/2b_0\aQ).
\ee
On the other hand, the function $\exp(-const/\alpha)$
with a positive $const$ is a classical example from 
the math courses of a function which has a trivial 
Taylor expansion at $\alpha=0$:
\be
\exp(-const/\alpha)|_{\alpha=0}~=~
0+0\cdot\alpha+0\cdot\alpha^2+...
\ee
since the function itself and all its derivatives vanish at $\alpha=0$.
Thus, this function, being a non-zero, vanishes identically
as a perturbative expansion, which is the expansion at $\alpha=0$. 

The interest in power-like corrections originates from various sources:

({\it i}) Nowadays, one may say that the perturbative QCD is trivially correct
and, if it were possible, the best thing would be just to
subtract it out and proceed to non-perturbative pieces. In
particular, one may say that the physics of the confinement is
encoded in the power corrections, not the perturbation theory.

({\it ii}) In some cases, the accuracy of theoretical fits to
experimental data require for an account of the power corrections. 
For example, error bars on measured values of $\alpha_s$ 
are affected by the power corrections.

({\it iii}) More pragmatically, one could say that we are brought
to consider the power corrections by the logic of 
the development in the field.

If it is at all possible to distinguish between these motivations,
we will belong rather to the first line. Namely, we will assume,
explicitly or implicitly,
that the onset of the power corrections at some so to say moderately large
$Q^2$ signifies new physical phenomena. At the end, we shall see whether
we are in fact justified to assume so.

There is another important aspect of the power corrections. While
calculating the perturbative expansions is a well defined procedure
in QCD, at least as a matter of principle, the definition of
the non-perturbative terms is close to saying that these are unknown
terms, the rest of the amplitudes upon subtraction of the perturbative
contributions. In other words, working with the power corrections   
relies to a great extent on intuition and heuristic models.
Nevertheless, we shall be mentioning sometimes the ``standard picture''
of the nonperturbative physics. What could this mean if we a priori know
that no precise form of the nonperturbative fluctuations is assumed?
Still, there is a content to the notion of the standard picture. Namely,
the standard assumption is 
that the nonperturbative fields are soft. In other words, the
typical size of the nonperturbative fluctuations is of order $\La^{-1}$.
Later, we will challenge this picture to some extent.

Finally, let us mention that actually working with two infinite series
as indicated in (\ref{two}) would be awfully difficult in practice.
Thus, in reality one is always relying on a kind of a {\it truncated}
series like:
\be
<O>~=<O>_{parton~model}\left(1+a_1\aQ
+b_k(\La/Q)^k\right)\ee
where $k$ characterizes the leading power correction.
The assumption behind this truncation is that
the power corrections are somehow {\it enhanced} numerically.
There is no proof of this assumption but it is an indispensable
ingredient of any phenomenology based on the power corrections.
Historically, this assumption worked very well in case of the so called
QCD sum rules (Shifman et. al, 1979). It might well fail, however, 
in other cases. If there exist high-precision data one may try to
check this assumptions varying the number of terms in
the perturbative expansion kept explicitly and watching how the fitted
value for the power corrections depends on this. So far, this
procedure was implemented in the most careful way in (Kataev et.al, 1998)
in case of DIS.

\section*{Outline of the lectures.}

The power corrections are a hot subject in QCD. In particular
at this School there will be another course, given by Yu. L. Dokshitzer
also devoted to a great extent to the power corrections. 
To avoid overlap, this lectures will consist of two parts.
The first one is about power corrections within 
the approach based on the operator product
expansion (OPE) which goes back to QCD sum rules and is about 20 years old.
The second part, to the contrary, describes an unconventional
(and hypothetical)
source of power corrections, that  is short strings. This part 
is based mostly on the original work  (Gubarev et. al., 1998).

It is worth adding that the unification of these two parts,
separated by almost 20 years in terms of the original papers, 
under the same 
title is not artificial at all. The point is that the power corrections
associated with the short strings hopefully solve some 
outstanding problems left
over from the OPE-based approach.  

\section{POWER CORRECTIONS AND SOFT VACUUM FIELDS.}

\section*{Correlation functions.}

It is  clear that if we take the limit $Q^2\to \infty$ literally
we would be left with parton model, with no corrections whatsoever.
Moreover, the power-like corrections would die fast.
Thus, our general strategy will be to start with large $Q^2$
so that $\aQ$ is small numerically. This is needed to have some control
over theoretical calculation since QCD is simple only
at short distances. However, 
then we will move down in $Q^2$ until reach so to say
moderate $Q^2$ where the corrections become sizable. This is most
interesting region for us. Indeed, we are still able to sort
out various corrections since they are small compared to unity.
On the other hand, we may hope to distinguish between various mechanisms
of breaking the asymptotic freedom.
In language of distances, we will start with $r\to 0$ and then proceed to
``moderate'' $r$. Everywhere we understand that generically, $r\sim 1/Q$.

For the sake of definiteness we will concentrate on correlation functions
$\Pi_j(Q^2)$
and quark-antiquark potential $V(r)$. Let us introduce these quantities
in more detail.

At first sight, it would be most natural for QCD studies to 
consider hadrons themselves.
However, observing hadrons we would not find much quarks at short distances
since they are predominantly at a characteristic distances of order
$\La^{-1}$. And physics at such distances is governed
by a large $\alpha_s$ where we do not have reliable theory.
To ensure that quarks do not fly away
one has to resort therefore to an external source of quarks such
as electromagnetic current and consider unphysical kinematics
with space-like total momentum of quarks $q$, $-q^2\equiv Q^2\gg \La^2$. Then
according to the uncertainty principle quarks can exist for time
of order
\be
\tau~\sim~{1\over \sqrt{Q^2}}\label{time}
\ee
which is small if $Q^2$ is large.
For consistency, after such time the quarks are to be absorbed by 
another
current, see Fig. 1.
\begin{figure}[h]
 \begin{center}
    \leavevmode
    \epsfig{figure=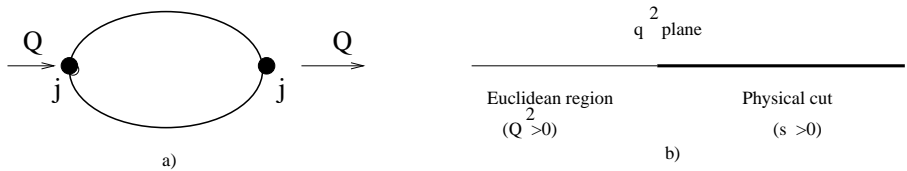,width=12cm}
    \caption{
a) Correlator of currents in the parton model approximation.
     {b) $q^2$ plane.} }
    \label{fig:fig1}
  \end{center}
\end{figure}

In the field theoretical language, we are considering in fact
a correlation function $\Pi_j (Q^2)$:
\be
\Pi_j (Q^2)~=~i\int d^4x~e^{iqx}~\langle 0| \; T\{j(x), j(0)\} \; |0\rangle,
~~~~Q^2\equiv-q^2 \label{correlator}
\ee
where the current $j$ may have various quantum numbers, like spin, isospin
and for simplicity we do not indicate these quantum numbers, i.e. suppress
the Lorenz indices and so on.

The basic theoretical ingredient is that $\Pi(Q^2)$ at large $Q^2$
can be calculated in the parton model approximation:
\be
\lim_{Q^2\to\infty}\Pi_j(Q^2)~=~\Pi_j(Q^2)_{\rm parton~model}
\ee
On the other hand, by using
dispersion relations
$\Pi_j (Q^2)$ can be expressed in terms of the absorptive part
which is non-vanishing only for time-like total momentum, $q^2>0$:
\be
\Pi_j(Q^2)~=~{1\over \pi}\int{Im\Pi_j(s)\over s+Q^2}ds\label{dre}
\ee
The imaginary part is directly observable, provided that
the current $j$ is a physical one. In particular,
in case of the electromagnetic current, $j=j_{el}$ the imaginary
part in Eq. (\ref{dre}) is proportional to the
total cross section of $e^{+}e^{-}$-annihilation into hadrons:
\be
Im\Pi_{j_{el}}(s)~=~const~{\sigma_{tot}(e^{+}e^{-}\rightarrow hadrons)
\over \sigma (e^+e^-\rightarrow\mu^+\mu^-)}
\label{st}
\ee
Upon substitution of (\ref{st}), the Eq. (\ref{dre})
becomes a sum rule. Indeed, $\Pi_{j_{el}}(Q^2)$ is calculable
and the same as for
free particles plus small radiative corrections, while $Im\Pi (s)$
is observable.

There is one more technical point to be mentioned.
The dispersion relations (\ref{dre}) suffer in most cases from 
ultraviolet divergences which could be eliminated at a price of
subtractions. But then there would appear
an arbitrary polynomial.
There is, however, another way to deal with this problem
(Shifman et.al., 1979). Namely  
introduce
\be
\Pi_j (M^2)~\equiv~{ Q^{2n}\over (n-1)!}
\left({-d\over dQ^2}\right)^n\Pi_j (Q^2)\label{di}
\ee
in the limit where both $Q^2$ and $n$ tend to infinity so that their
ratio $M^2\equiv Q^2/n$ remains finite.
Then it is easy to see that the weight function in the dispersion
relation becomes:
\be
{1\over s+Q^2}~\rightarrow~{1\over M^2}\exp(~-s/M^2),
\ee
while all the subtraction constants are removed by the differentiation
(\ref{di}).

Let us give a realistic example of the sum rule in the $\rho$-meson
channel:
\be
\int R_{I=1}(s)\exp(-s/M^2)ds~\approx~{3\over 2}M^2\left(
1+{\aM\over \pi}\right),
\label{rho}\ee
where $M^2$ is considered to be large enough so that
$\aM/\pi$ is small. Moreover $R_{I=1}(s)$ is the total cross section
of $e^+e^-$ annihilation into hadrons with isotopic spin I=1
in units of the standard cross section $\sigma(e^+e^-\rightarrow\mu^+\mu^-)$.

It is worth noting that to apply the technique considered
it suffices to ensure that the time which quarks exist
is small indeed. Apart from imposing the condition that
$Q^2$ is large (see Eq. (\ref{time})) there exist other possibilities.
In particular, as far as production of heavy quarks is concerned
one can consider $Q^2=0$ since in that case 
(Shifman et. al, 1976) $$\tau\sim 1/m_H~.$$
Therefore, in case of heavy quarks it is convenient to
consider $Q^2=0$ and get rid of possible ultraviolet (UV) divergences by
differentiating the dispersion relations
(\ref{dre}) with respect to $Q^2$ at $Q^2=0$. In this way
one comes to the sum rules (Novikov et. al., 1977):
\be
\int{R_c(s)~ds\over s^{n+1}}\approx~
{A_n\over (4m_c^2)^n}\left(1+B_n\alpha_s(m_c^2)\right),\label{psi}
\ee
where $R_c$ is the contribution of the current of
the charmed quarks into the ratio $R(s)$ and 
the integer number $n$ corresponds to the $n$-th derivative from
(\ref{dre}) while $m_c$ is the mass of the charmed quark
normalized off-mass shell at $p^2=0$.
Moreover, $A_n,B_n$ are calculable numbers:
\beqn
A_n~=~{3\over 4\pi^2}{2^n(n+1)(n-1)!\over (2n+3)!!},\nonumber \\
B_n~=~{4\sqrt{\pi}\Gamma(n+3/2)[1-1/(3n+3)]\over 
3\Gamma(n+1)[1-1/(2n+3)]}-{\pi\over 2}+{3\over 4\pi}-\\
-{2\over3\sqrt{\pi}}\left({\pi\over 2}-{3\over 4\pi}\right)
{\Gamma(n+3/2)[1-2/(3n+6)]\over
\Gamma(n+2)[1-1/(2n+3)]}-{4n\; \ln 2 \over \pi}\label{ch}\nonumber
\eeqn

Note that the integral over $R(s)$ is contributed both by resonances, 
like $J/\Psi$ and by continuum cross section of 
production of particles with open charm.
The bound states are to be included since their contribution is of
zero order in the small $\aM$. Indeed, 
the properties of the resonances are governed by $\alpha_s\sim 1$.
Moreover it turns out that there exist such $M^2$ (or $n$ in case
of heavy quarks) for which the resonances dominate the integral over
the physical cross section while the theoretical, partonic part
is still calculable reliably since $\aM$ is small.

It is worth noting that direct experimental information on the 
correlation functions is available only in a limited number of cases
since only very few currents, like the electromagnetic current
is directly observable. However, the lattice simulations 
(Chen et. al., 1993) allow to measure $\Pi_j(x)$ for a much wider class of
currents. These measurements refer to the Euclidean space directly
since the lattice corresponds, of course, to Euclidean $x$.

{\it To summarize:} there are quite a few
correlation functions known in Euclidean domain
either from combined use of dispersion relations and 
experimental data or
from the lattice simulations. The correlation functions
are most suitable for studies of the power corrections.

\section*{ The meaning of the OPE. Gluon Condensate.}

In this section we will describe power corrections as
they arise within the operator product expansion (OPE).
The presentation goes back to the QCD sum rules
(Shifman et. al., 1979). More detailed reviews can be found in
(Reinders et. al., 1985), (Narison, 1989).

The central physical question which brought about the power corrections
was as follows. Imagine that we study the sum rules (\ref{rho})
at very large $M^2$ and then go down with $M^2$. At some low
$M^2$ the dispersive integral over $s$ would be dominated by the 
$\rho$-meson to such extent that it cannot be matched by the quark
predictions. Then the question is:
``Who stops the Asymptotic Freedom at some critical $M^2$?''

At first sight, the answer is almost trivial: the growth
of the coupling at low momenta. It would be a common answer.
However, if one addresses the problem in concrete,
there arise doubts in the validity of this answer.
For example, it is quite a common
theoretical guess that the splitting between
the vector mesons does not depend on flavor, say:
\be
m(\rho')-m(\rho)~\approx~m(\psi')-m(J/\psi)\label{simplee},
\ee
which is true experimentally.
This simple-looking observation represents, 
however, a serious challenge to the
wisdom that it is the growth
of the effective coupling (\ref{coupling})
that stops the AF at moderate mass scales.
Indeed, if expressed in terms of an invariant quantity, $s$,
Eq. (\ref{simplee}) implies that the $J/\psi$ is dual to a much
larger interval of $s$ than the $\rho$
because the c-quark is heavy. In other words, the AF is violated
in the $\rho$ channel much later than in the charmonium channel,
if we start from very large $M^2$ downwards.
A direct numerical analysis of the sum rules (\ref{rho}), 
(\ref{psi}) confirms
this expectation. However, the coupling should run as a function
of an invariant quantity and is flavor blind.

More generally, one feels that knowing the perturbation theory alone
is not enough to plunge into the resonance physics. 
Therefore, let us assume for the moment that it
is soft nonperturbative fields in the vacuum
which are eventually responsible for the confinement.
At first sight, having this idea does not help much
since very little is known on the precise nature of these
nonperturbative fields.
The way out of this difficulty will be not to try to calculate the
nonperturbative fields but describe
them instead phenomenologically in terms of a few parameters.

One can understand the trick by considering the same Feynman
graphs for the $\Pi_j(Q^2)$. Let us begin with the simplest graph of 
Fig. 2. As was explained above (see Eq. (\ref{time})), by taking $Q^2$ large
we ensure that the correlation function is determined by
short distances. Let us now estimate to which accuracy this
assertion is correct.
Consider to this end in more detail the route of the 
large $Q$, which is brought in by the current.
The typical case is when the both quark lines in Fig. 1
carry momentum of order $Q$. ``Typical'' means favored by the phase space.
But it is not forbidden that almost the whole momentum is carried
by one line while the other quark line is soft. Then we get 
a problem since the soft line cannot be reliably described by the 
perturbative propagator because at large distances the perturbation
theory is modified strongly.

Instead of making guesses, how the quark propagator 
looks like in the infrared (IR)
we reserve for an unknown number for this propagator.
The central point is of course that it is not an unknown function
but rather a number. To see that it is indeed true, 
prepare the graph as is shown in Fig. 2. 
\begin{figure}[h]
 \begin{center}
    \leavevmode
    \epsfig{figure=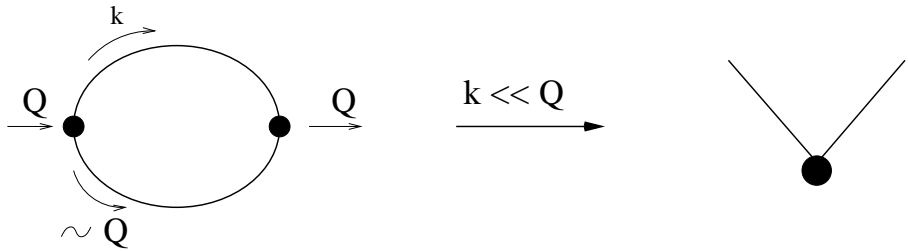,width=12cm}
    \caption{
If one of the lines is soft, one substitutes instead the perturbative
propagator the corresponding vacuum matrix elements of the fermionic operator.
}
    \label{fig:fig2}
  \end{center}
\end{figure}
Here the black blob in the right-hand side
denotes the hard quark line and currents. Along this line, since it
is hard, all the distances are small, $x\sim 1/Q$ and everything is
calculable. On the other hand, the soft line can travel
distances of order $1/k\sim \La^{-1}$ which are large on the scale of
$1/Q$. However, it is constrained to begin and end up at the same point.
We can substitute therefore the soft line by an unknown matrix element.  
This matrix element is proportional to the so called quark condensate,
$
<0|\bar{q}q|0>~\neq~0
$,
which is famous for the fact that its non-vanishing value signals the
spontaneous breaking of the chiral symmetry.
If the current $j$ is constructed on the light quarks, then it is easy
to see that in fact the quark condensate is multiplied by the quark mass
$m_q$ and the contribution of the IR region is additionally suppressed.
Namely it is of the order
\be
{\delta\Pi_j(Q^2)\over \Pi_j(Q^2)}~\sim~{m_q<0|\bar{q}q|0>\over Q^4}
\label{cond}
.\ee 
Notice the factor $Q^{-4}$ which we get on pure
dimensional grounds. This suppression is the price we pay for 
enforcing one of the lines to be soft.

The graph in Fig. 2 is very important,
however, if one of the quarks is heavy.
Then we replace $m_q$ in Eq. (\ref{cond}) by $m_H$ while $q$ still stands
for a light quark.
This is our first example of a nontrivial interplay
of perturbative and nonperturbative calculations.
Indeed, on one hand, we consider perturbative graph. The perturbation
theory by itself respects the symmetries of the Lagrangian and
$<0|\bar{q}q|0>_{pert}\sim m_q$ and is negligible. However, in the general
treatment of the infrared sensitive contributions which we pursue now
the quark condensate $<\bar{q}q>$ is just a phenomenological number
which is in fact does not go to zero for $m_q\to 0$.

Coming back to consider currents constructed on the light quarks,
the contribution (\ref{cond}) is small because $m_q$ is small
and we have to go to the next order in perturbation theory
to uncover IR sensitivity of the Feynman graphs at large $Q$. 
Thus, turn to the graph with
a gluon exchange.
Assuming that the momentum flowing through the gluon line is in fact 
of order $\La$ we prepare the graph as indicated in Fig. 3.
\begin{figure}[h]
  \begin{center}
    \leavevmode
    \epsfig{figure=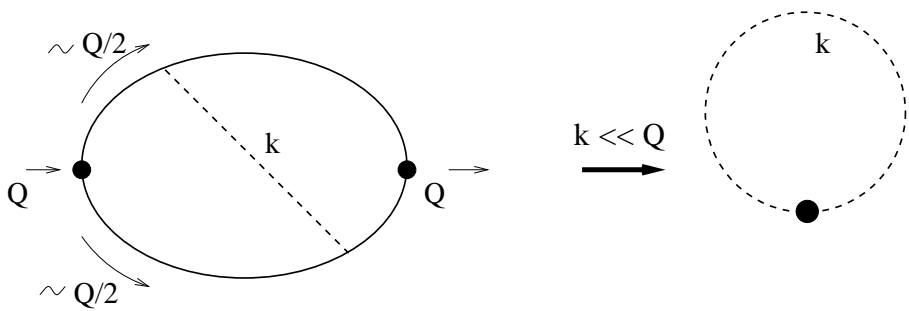,width=12cm}
    \caption{Treatment of one-gluon exchange graphs according to the OPE
in case when the gluon line is soft. The black blob corresponds
to the hard quark lines and the currents. All the points here are close to
each other. Instead of the soft gluon line, denoted by a dashed line,
one substitutes a phenomenological matrix element.}
    \label{fig:fig3}
  \end{center}
\end{figure}
Again, there is no reason any longer to use the
perturbative expression for the gluon line since it is soft and
modified strongly by the confinement.
The quark lines propagating short distances become a receiver of
long wave gluon fields in the vacuum.
The receiver is well understood because of the asymptotic freedom.
The intensity of the gluon fields is measured. It is characterized
by the so called gluon condensate:
\be
\langle 0|\alpha_s(G_{\mu\nu}^a)^2|0\rangle~\equiv~
\langle 0|\alpha_s\left( (\vec{H}^a)^2 -(\vec{E}^a)^2\right) |0\rangle
~\neq~0\label{gc}\ee
where $\vec{H}^a, \vec{E}^a$ are color magnetic and electric fields
and $G_{\mu\nu}^a$ is the gluonic field strength tensor.

The question may arise, how do we know that it is the matrix element
(\ref{gc}) that arises in a straightforward evaluation of the
IR sensitive part of the one-gluon exchange graph.
The answer to this question is not difficult: 
this is the simplest expression compatible with 
the Lorentz and gauge invariance. In other words, the $(G_{\mu\nu}^a)^2$ is the
simplest operator which is Lorentz and color singlet. It has dimension d=4
implying that the IR sensitive contributions are suppressed at large
$Q^2$ as $Q^{-4}$.

What is described here in words, 
in its generality has an adequate formulation
in terms of the Wilson operator expansion (Wilson, 1969).
A systematic treatment of the power corrections within
the Wilson OPE was given in (Shifman et. al., 1979).
In particular, it allows to evaluate 
the contribution of the gluon condensate (\ref{gc})
to correlation functions corresponding to various currents $j$. 

One may still wonder, how it could be possible 
to go beyond the perturbation theory
through a simple preparation
of perturbative graphs. 
To answer this question, let us try to evaluate the 
gluon condensate in perturbation theory. To lowest order it is given by
a simple one-loop graph:
\be
\langle 0|\alpha_s(G^a_{\mu\nu})^2|0\rangle~\approx~
{8\cdot 3\over (2\pi)^4}
\int d^4k{k^2\over k^2}\label{divergent}
,\ee 
where we did not cancel immediately $k^2$ in the denominator 
and 
numerator of the integrand to emphasize that the factor $k^{-2}$ is the gluon
propagator while the factor $k^2$ upstairs is associated with the
derivatives in the vertex. Also, we assumed that there are eight color gluons,
as in the realistic case. From the first glance at 
the expression (\ref{divergent})
it is clear that it diverges in the ultraviolet as $\Lambda_{UV}^4$.
However, in all the applications we assume instead that
\be
\langle 0|\alpha_s(G^a_{\mu\nu})^2|0\rangle~\equiv~c_G\La^4\label{postulate}
\ee
where the constant $c_G$ is to be found from the fitting procedure.
And there is a rational behind this madness.
The point is that the UV divergent
part of the integral (\ref{postulate})
corresponds the ordinary perturbative contribution.
(Of course, there is no real UV divergence 
in the original Feynman graph (see Fig. 3) since at $k\sim Q$ the account
for
other lines in the graph provides with a kind of
form factor so that the final result is finite.) Thus, we include
only the IR contribution into the gluon condensate.
Generally speaking, we should have subtracted then the perturbative IR
contribution, included now into the matrix element of $(G^a_{\mu\nu})^2$
from the ordinary perturbation theory.
This is to avoid a double counting,
as was emphasized by David (1984) and Mueller (1985).
There are no difficulties of principle to
deal with the problem (see, e.g., Shifman et al (1979),
Novikov et.al. (1985)).
However the whole approach is becoming most interesting
from the phenomenological point of view if the nonperturbative
contribution is {\it enhanced} numerically. And this is the assumption
which is commonly made at this point. Thus, the OPE allows to 
introduce in a consistent fashion the notion
of the enhancement of the IR sensitive contributions
to Feynman graphs.   

Coming back to the problem of isolating soft lines
in perturbative graphs,
the same one-gluon exchange graph can be characterized by
another route of the large $Q$ flow, see Fig. 4.
\begin{figure}[h]
 \begin{center}
    \leavevmode
    \epsfig{figure=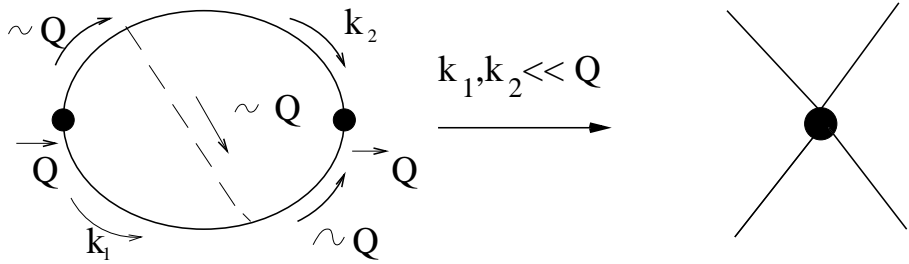,width=12cm}
    \caption{Space-time picture corresponding to the Wilson operator product 
expansion in case that there are two soft quark lines.}    
\label{fig:fig4}
  \end{center}
\end{figure}
Here we have two soft quark lines which means in turn
that the corresponding correction to the right-hand side of the sum rules
(\ref{rho}) is of the type:
\be
{<0|\bar{q}Oq\bar{q}Oq|0>\over Q^6}
,\ee
where $O$ is constructed on the Dirac matrices $\gamma_{\mu}$
in the spinor space 
and on the Gell-Mann matrices $\lambda^a$ in the color space.
Usually one assumes factorization of the 4-quark matrix element
reducing it to the quark condensate mentioned above. 
This time, however, the
condensate is not suppressed by power of the (practically vanishing) 
light quark masses. This is another example how the nonperturbative
phenomenon of the spontaneous breaking of the chiral symmetry finds
its way into the OPE based on analysis of the perturbative graphs.

With account of the quark and gluon condensates the sum rules 
in the $\rho$-channel become:
\beqn\label{qcd}
\int R_{\rho}(s)\exp(-s/M^2)ds~\approx~{3\over 2}M^2
\times(
1+{\alpha_s(M^2)\over \pi}+\\ \nonumber
{\pi\over 3}{\langle 0|\alpha_s(G^a_{\mu\nu})^2|0\rangle \over M^4}
-{32\cdot 14\pi^3\over 81}{|\langle 0|
\alpha_s^{1/2}\bar{q}q|0\rangle |^2\over M^6}+...)
\eeqn
where the ellipses stand for
perturbative and power corrections
of higher order.
In case of the charmonium sum rules (see (\ref{ch}))
there also appears a term proportional to the gluon condensate.

Eqs. (\ref{qcd}) are the QCD sum rules (Shifman et. al., 1979).
The first check was to see whether the gluon condensate
explains the difference in the duality intervals in the
$\rho$- and $J/\psi$- channels (see above). It does explain
this difference naturally.
Also in many other cases,
the QCD sum rules turned to be a very
straightforward and successful
tool for orientation in the hadronic world.
For our review it is most relevant that the sum rules
provide with ample confirmation of the idea that
it is the soft vacuum fields that are responsible for the breaking of 
the asymptotic freedom at moderate $M^2$ and
signal in this way the formation of the resonances.
One can characterize the power corrections by $M_{crit}^2$
which is defined as the value of $M^2$ where the power corrections 
become, say, 10\% of the partonic contribution. 
Then the typical $M_{crit}^2$ associated with the quark and gluon
condensates is
\be
M^2_{crit}~\approx~0.6~ \mbox{GeV}^2\label{mcr}
\ee
and its dependence on the channel for currents constructed on
light quarks is not strong. 
The meaning of the $M^2_{crit}$ is that the power corrections blow
up at smaller $M^2$. Clearly, the value (\ref{mcr}) resembles
the $\rho$-meson mass squared.

{\it To summarize:} the use of OPE allows to parameterize infrared
sensitive contributions to the Feynman graphs in a systematic way
in terms of vacuum expectation values.
While the coefficients in front of the matrix elements are calculable
as an expansion in $\aQ$, the
matrix elements encode the information on nonperturbative phenomena.

\section*{The heavy quark potential.}

Another useful quantity to consider in connection with the power corrections
is the heavy quark potential at short distances, 
for a review and further references see (Akhoury and Zakharov, 1998).

Let us first notice that although the static potential between heavy
quarks is obviously a fundamental quantity, its definition in
non-Abelian theories is not straightforward at all.
The point is that we are interested in case when the quarks are
in a color-singlet state. On the other hand, the  
gauge invariance allows to
rotate quarks in the color space locally and, at first sight,
it is just impossible to single out a color singlet state.
The way out is to consider the so called Wilson loop,
that is the average of an operator constructed on the
gauge field potential $A_{\mu}^a$ along a closed path $C$.
Moreover, it is convenient to choose the contour $C$ as
a stretched rectangle $C=r\times T$ with large $T$.
Then the potential $V(r)$ is:
\be
V(r)~=~-\lim_{T\to\infty}{1\over iT}\ln\langle
Tr~P~\exp\left(ig\oint_Cdx^{\mu}A_{\mu}^aT^a\right)\rangle.
\ee
This definition is gauge invariant and refers to the Euclidean space.
The main source of knowledge of $V(r)$ are the lattice simulations
(see Bali et al, (1995); Bali, (1999)). 

The most famous property of the potential $V(r)$ is that it 
grows linearly at large distances $r$:
\be
\lim_{r\to\infty}V(r)~=~\sigma_{\infty}r
.\ee
Note that here we imply that the dynamical light quarks are not taken
into account and consider the interaction of infinitely heavy external
quarks embedded into the vacuum of pure gluodynamics.

We are mainly interested in the potential $V(r)$ at {\it short} distances.
Let us begin our analysis with a trivial Yukawa 
attractive potential and expand it in powers of $(\lambda r)$
at small r:
\be
V_{Yu}(r)~=~-{C\over r}e^{-\lambda r}~=~-{C\over r} +C\lambda-
{Cr\lambda^2\over 2}+
{Cr^2\lambda^3\over 6}+...\label{yukawa}
~~,\ee
where $C$ is a positive constant.
This simple equation carries an important message,
namely, that the physics behind
the odd and even powers of the mass $\lambda$ in the expansion (\ref{yukawa})
is different. Indeed, naively we should have
had only powers of $\lambda^2$ since $\lambda^2$ 
is the only mass parameter entering the
Lagrangian of a massive boson. Thus, our idea could be that we start
with a Coulomb like potential at small $r$ and develop a perturbative
expansion in $\lambda^2r^2$. Eq. (\ref{yukawa}) implies that such an approach
would fail because of infrared divergences.
This can be readily checked explicitly of course. But it suffices to notice 
that the odd powers in (\ref{yukawa}) could arise only from infrared
divergences which are cut off at distances of order $1/\lambda$.

Appearance of terms non-analytical in $\lambda^2$ in (\ref{yukawa})
allows to make 
important conclusions about 
the heavy quark potential in QCD.
Generally, 
if a physical observable is sensitive to large distances 
at level of a certain power correction
then it cannot be evaluated to such accuracy
because the effective coupling is large. Rather, one should reserve for
the strength of the correction as a phenomenological parameter.  
In our case, we can conclude from Eq. (\ref{yukawa})
that the quark-anti-quark potential in QCD can be parameterized as:
\be
V(r)~=~-{C_{-1}\over r}+C_0\La+C_2r^2\La^3+...\label{qq}
\ee
where $C_{-1}$ is calculable perturbatively as an expansion in
$\alpha_s(r)$ while $C_{0,2}$ account phenomenologically 
for the contribution of large distances. Moreover, $C_0$ can be included
into definition of the (heavy) quark masses and the message is that the power
corrections to the potential at short distances start with $r^2$
terms (Balitsky (1986). Dosch and Simonov(1988)).
A salient feature of Eq. (\ref{qq}) is absence of a linear in $r$ term
at {\it short} distances. The proof was in two steps. First, we learn
from (\ref{yukawa}) that the linear correction 
to the potential is not contaminated by large-distance effects and,
then, conclude that these corrections vanish since 
the only parameter of dimension d=2 is
the gluon mass squared and it vanishes in QCD. 
Thus, presence of a linear term
at short distances in the quark potential would signal 
new non-perturbative physics at short distances.

The crucial assumption behind Eq. (\ref{qq})
is that typical size of nonperturbative fluctuations is
of order $\La^{-1}$. 
This connection can readily be visualized in the abelian case.
Namely, begin with an identity for the potential:
\be
V(r)~=~{1\over 4\pi}\int d^3r'{\bf E}_1({\bf r}')\cdot{\bf E}_2({\bf r+r'}),
\label{space}\ee
where ${\bf E}_1,{\bf E}_2$ are electric fields associated with each
charge. Eq. (\ref{space}) is convenient to account for the effect of the
change in the electric fields at large distances. 
Consider an example of two charges of opposite signs in a cavity of size $R$.
The electric field of the charges 
in empty space is that of a dipole at large 
distances, ${\bf E}^2\sim \alpha r^2/r^{'6}$.
Now, because of the cavity the field 
is changed at $r^{'}\sim R$ and the corresponding feedback to
the potential $V(r)$ is of order:
\be
\delta V(r)~\sim~\alpha r^2
\int_R^{\infty}{d^3r'\over (r')^6}~\sim ~{\alpha r^2\over R^3}
\ee
which is in agreement with the correction to the static potential discussed
above. 
 
After, hopefully, explaining thoroughly that the correction to the
potential $V(r)$ at short distances is of order $r^2$ in the standard picture
we are going to introduce a new notion, that is retardation 
effects (Casimir and Polder, 1948).
Discussion of this point is a kind of deviation from 
the course to our immediate goals. We are in fact interested in 
the static potential at short distances. 
And then there are no retardation effects, of course.
However, historically atom-like systems of heavy quarks were discussed 
at most and in this case the retardation effects may be important.
This is the only reason to include discussion of the retardation effects into
this review. 

Thus, Casimir and Polder did discuss the interaction of atoms with 
nonperturbative long-wave fields in QED about 25 years 
before the advent of QCD. What kind of nonperturbative fields 
did they consider?
At first sight at least, there no nonperturbative fields in QED.
They considered atom in a cavity. Then zero-point fluctuations
are modified by the presence of the cavity at frequencies of order
$\omega\sim R^{-1}$ where $R$ is the cavity size. If
in the QCD case, we substitute the cavity of size $R$
by a nonperturbative field of the size $\La^{-1}$,
the problems turn similar!

Of course, here we present only a very sketchy view of the paper by 
Casimir and Polder. One starts with the dipole interaction (see 
also above):
\be
H_{int}~=~-e{\bf d \cdot E}.
\ee
The shift in energy levels of atoms arises then in second order in
this interaction:
\be
\delta E_n~\sim~\sum_k V_{nk}(E_n-E_k+\omega_{char})^{-1}V_{kn}
\label{denominator},\ee
where the characteristic frequency 
of the nonperturbative fields
is of order $\omega_{char}\sim 1/R$.
The result for the energy shift $\delta E_n$ depends crucially 
on the relative magnitude of $\omega_{char}$ and
$E_n\sim m\alpha^2\sim\alpha/a_B$ where $a_B$ is the Bohr radius.
Namely, if $\omega_{char}\gg E_n$, then
\be
\delta E_0~\sim~\alpha{\bf d}^2{\bf E}^2 R~\sim~\alpha{a_B^2\over R^3},
~~~R\ll a_B/\alpha\label{shift}
\ee
since ${\bf d}^2\sim a_B^2$ and ${\bf E}^2\sim R^{-4}$.

The shift (\ref{shift}) corresponds to our estimate given above in
the language of the classical electrodynamics. In this case, 
the potential picture applies to the evaluation of the energy shifts.

On the other hand, if $R\gg a_B/\alpha$ then $\omega_{char}$ in the energy
denominator in (\ref{denominator}) can be neglected and 
\be
\delta E_0~\sim~\alpha {\bf d}^2{\bf E}^2{a_B\over \alpha}~
\sim~{a_B^3\over R^4},~~~
R\gg a_B/\alpha ,\label{violation}\ee
where we used 
the estimates of ${\bf E}^2, {\bf d}^2$ given above and substituted
$E_n\sim \alpha/a_B$.

The estimate (\ref{violation})
is in clear violation of the potential picture. Moreover, 
Eq. (\ref{violation}) could be interpreted
by saying that if the distance $a_B$ between the particles   
is much smaller than $\alpha R$
then the  $r^2$ potential is replaced by a $r^3$ potential 
\footnote{But this is true only as far
as rough estimates are concerned. Rigorously speaking, there is no potential 
whatsoever 
corresponding to the shifts obtained in this way. This
was emphasized much later, in connection with QCD,
by Voloshin (1979) and Leutwyler (1981).}.
Note also that the emergence of the scale $R\sim a_B/\alpha$ 
in the equations above can be understood as an effect of retardation.
Indeed, the time needed to communicate with the distances of order
of the cavity size $R$ can be called the retardation time, $T_{ret}\sim R$.
For the potential picture to be valid, $T_{ret}$ is to be smaller than 
the revolution time which is of order, $T_{rev}~\sim~a_B/v~\sim
a_B/\alpha$. The potential picture becomes distorted once 
$T_{ret}\approx T_{rev}$.

In QCD case one considers (see Voloshin (1979), Leutwyler (1981))
atom-like systems of heavy quarks $Q$ such that
\be
\alpha_s(M_H)\cdot M_H~\gg~\La\label{condition}
\ee
where $M_H$ is the mass of the heavy quark and the condition
(\ref{condition}) is that the Bohr radius is much smaller than 
$\La^{-1}$. Again, the dipole interaction is relevant to describe
interaction with soft gluon fields:
\be
H_{int}~=~-\sqrt{\alpha_s}(t_1^a-t_2^a){\bf d \cdot E}^a,
\ee
where $t_i^a~(i=1,2)$ are generators of the color group and refer to
the quark $Q$ and anti-quark $\bar{Q}$ in the quarkonium
while ${\bf E}^a$ denotes the soft gluonic field in the vacuum.

Moreover, one assumes that $\omega_{char}\sim\La^{-1}$. As for the
intensity of the gluonic fields, it is characterized again
by the gluon condensate $<\alpha_s(G_{\mu\nu}^a)^2>$.
The resulting shift of the energy levels of the $S$-states depends
strongly on the principal quantum number $n$:
\be
{\delta E_{nl}\over M_H}~=~n^6{\pi <\alpha_s(G_{\mu\nu}^a)^2>
\over (M_HC_F\alpha_s)^4}\epsilon_{nl}\label{vl}
,\ee
where the numbers $\epsilon_{nl}$ depend on the quantum numbers
of the level, $n,l$, and are of order unity, e.g., 
$\epsilon_{10}\approx 1.5$.
The growth with $n$ is due to the growth of the size of the atom-like state.
However, to apply (\ref{vl})
one should assume that the retardation effects are already very important.
For a concrete value of the heavy mass $M_H$ the balance between
these requirements can be quite delicate and we refer
the reader to a review by Yndurain (1998) for details and references.

{\it To summarize:}
according 
to the standard QCD there is no linear correction to the Coulomb potential
at short distances. This is in variance with naive unification
of the linear potential (dominating at large distances) and Coulomb-like
potential (dominating at small distances).

Moreover, because of the retardation effects the effect of the 
leading $r^2$ correction in $V(r)$
would be washed out in heavy quarkonium and the nonperturbative
corrections
to the Coulombic potential at short
distances would start with even smaller terms of order $r^3$.

\section*{Other techniques: infrared renormalons, infinitesimal gluon mass.}
 
Notice that while discussing $V(r)$ we did not 
resort to the OPE. It is not by chance of course. The point is that 
no matter how simple the notion of the potential $V(r)$ 
could appear, the OPE cannot be applied directly to evaluate $V(r)$.
The reason is that we consider the static potential
which means, potential energy averaged over large period of time.
Thus, although we consider small spatial distances $r$, 
the potential $V(r)$ is sensitive to large intervals
in the time directions.

It is not the only case when the OPE does not apply
of course. Moreover, as was  mentioned above,
the OPE applies actually in the Euclidean space and only in some
cases, when the analytical properties are simple, the results
can be continued to the Minkowski space.

Thus, it is desirable to develop
techniques to evaluate the power corrections directly
in the Minkowski space. And there exist such techniques.
In particular, we will discuss here infrared renormalons
and infinitesimal gluon mass. We do not have time to cover the topics in
detail. Instead we will illustrate the technique by two examples,
rederiving the results which we already know.
A more detailed exposition and further references can be found in
reviews, see, e.g., Dokshitzer et. al. (1996), Akhoury and Zakharov (1997),
Zakharov (1998), Beneke (1998).

First, we will consider the power corrections to
the potential $V(r)$ by using the IR renormalon technique
(Aglietti and Ligeti, (1995); Akhoury and Zakharov, (1998 )). 
Consider to this end the one gluon exchange potential
\be
V(r)~=~-C_F\int {d^3{\bf k}\over (2\pi)^3}4\pi\alpha_s({\bf k}^2)
{\exp(i{\bf k\cdot r})\over {\bf k}^2},\label{potential}
\ee
where $C_F$ is related to the color indices and $C_F=4/3$ 
in the realistic case of the $SU(3)$ color group.
Note also that running of the $\alpha_s({\bf k}^2)$ 
incorporates the leading logarithmic corrections.
To find the renormalon contribution we rewrite 
(\ref{coupling}) identically as
\be
\alpha_s({\bf k}^2)~=~\int d\sigma\left({k\over \La}\right)^{-2\sigma b_0}
.\label{sigma}\ee

Next, substitute (\ref{sigma}) into (\ref{potential})
and perform integration over directions of {\bf k}
to get
\be
V(r)~=~{2\over 3\pi^2}\int_0^{\infty}dk
{\sin(kr)\over kr}\left({k\over \La}\right)^{-2\sigma b_0}
.\ee
Moreover, since we are interested  in $V(r)$ at small $r$ we expand in $kr$:
\be
V(r)~=~-{2\over 3\pi^2}\int_0^{\infty}d\sigma\Lambda_{QCD}^{2\sigma b_0}
\int dk(k^{-2\sigma b_0}+k^{2-2\sigma b_0}r^2+...).
\ee
Now, integrating over $k$ introduces what is called renormalon poles
at
\be
\sigma_{pole}~=~{1\over 2}b_0,~{3\over 2}b_0~...
\ee
The next integration, that is over $\sigma$, becomes undefined
as a result of these poles. One can resort, say, to defining the integrals
as their principal values. But this prescription is openly arbitrary
and one is to reserve for an overall rescaling of the
pole contributions.
As a result, we get
as corrections to the Coulombic potential at short distances:
\be
\delta V(r)~=~c_0\La+c_2\La^3r^2+...\label{correction}
\ee
reproducing the result speculated upon above.
Since we expand in $kr$ and $k\sim\La^{-1}$,
Eq. (\ref{correction}) is valid at
$r\ll \La^{-1}$

Since we were able to determine the leading powers for the power
corrections by means of the renormalons, one could think that
the OPE is no better than the renormalons. It is, unfortunately, not
true. Namely, renormalons do not 
allow, generally speaking, for a systematic expansion
of the coefficients in front of the power corrections in
small coupling, like $\alpha_s({\bf k}^2)$. 
Generally speaking, {\it all orders of perturbative expansion collapse
to the same order contribution if projected onto the power corrections.}

Let us illustrate the point by the same example of the quark potential.
In fact three first orders of the perturbative expansion 
in front of $1/r$ were explicitly found, see Peter (1997):
\be
V({ k}^2)~=~-{4C_F\pi\alpha_{\overline{MS}}\over {k}^2}
\left(1+{\alpha_{\overline{MS}}\over \pi}(2.583-0.278n_f)+
{\alpha^2_{\overline{MS}}\over \pi^2}(39.650-4.147n_f+0.077n_f^2)\right)
\ee
Moreover, in the approximation of the one-loop $\beta$-function,
\be\as^2({{k}}^2)~=~{1\over 2b_0}\La{d\over  d\La}\as({k}^2).
\label{reduction}\ee
By differentiating $n$ times $\alpha_s$ with respect to $\ln\La$ 
we immediately find contribution of $(n+1)$ renormalon chains, 
associated with the $\as^{n+1}({{k}}^2)$ in the perturbative expansion.

Thus, we have a unique possibility to compare the 
contributions of one-, two- and three renormalons:
\be\delta V(r)~\approx~(const)\La^3r^2(1+1.1+6.0+..).\label{numbers}\ee
We see that there is no convergence in sight, even numerically
\footnote{Strictly speaking, we should have kept further terms in
the $\beta$-function since we consider a multi-loop effect. Thus,
the example is still rather illustrative than rigorous.}.

The general rule is that if perturbatively 
$$V(k^2)=(4\pi\as/k^2)f(\as)$$
then \be
\delta V(r)~=~v_0\La f^B(\as=1/2b_0)+v_2\La^3r^2f^B(\as=3/2b_0)\ee
where 
$v_{0,2}$ are constants and
$f^B$ is the Borel-improved expansion, i.e. the series with the
expansion coefficients $a_n^B=a_n/n!$.

Another technique which can be used directly
in the Minkowski space is 
the introduction
of a (fictitious) gluon mass $\lambda, \lambda\to 0$
and tracing terms non-analytical in $\lambda^2$.
In fact, we used this technique when analyzed
the Yukawa potential. Now we would like to emphasize its generality.
Moreover,
the idea is similar in fact to that underlying introduction of the
gluon condensate.

Indeed, let us consider again the gluon condensate 
perturbatively (see Eq.(\ref{divergent})) but this time
in case of a finite gluon mass
(see Fig. 5):
\be
\langle \alpha_s(G_{\mu\nu}^a)^2\rangle~=~{6\alpha_s\cdot 8\over (2\pi)^4}
\int {d^4k~ k^2\over k^2+\lambda^2} .\label{lgc}\ee
It diverges wildly in the ultraviolet, as discussed.
Let us define the gluon condensate,
however, as the non-analytical in
$\lambda^2$ part of the perturbative answer 
(Chetyrkin and Spiridonov, 1988):
\be
\langle 0|\alpha_s (G^a_{\mu\nu})^2|0\rangle~=
{3\alpha_s\over \pi^2}\int_0^{\infty} {k^4dk^2\over (k^2+\lambda^2)}
\equiv -{3\alpha_s\over \pi^2} \lambda^4\ln\lambda^2,\label{definition}
.\ee
Note that there is no difficulty in practice to pick up the non-analytical
in $\lambda^2$ terms. To this end, it is sufficient to differentiate
(\ref{lgc}) twice with respect to $\lambda^2$.

Moreover, to finally
get rid of the gluon mass (which is not a pleasant sight for
a theorist's eye) we make a replacement:
\be
\alpha_s\lambda^4\ln\lambda^2~\rightarrow~c_4\La^4\ee
where $c_4$ is an unknown coefficient \footnote{
The technique with $\lambda\neq 0$ is in fact close to
the technique based on 
the infrared renormalons, mentioned above.. 
In the context of the gluon
condensate, the infrared renormalons were
considered first in (David, 1984) and (Mueller, 1985).}.
The central point is that if we evaluate various correlation functions
(\ref{correlator}) and isolate the terms $\lambda^4\ln\lambda^2$
we would reproduce the sum rules (23). Indeed, as we explained
above the gluon condensate in the sum rules (23) parameterizes
the infrared sensitive part of the Feynman graphs associated with
a soft gluon line. Picking up terms non-analytical in $\lambda^2$
is the same good for this purpose since the non-analyticity in
$\lambda^2$ can obviously arise only from soft gluons, 
$k\sim \lambda$.

So far we have not got any new result, though.
The real advantage of introducing $\lambda\neq 0$ is that
calculations can be performed now in Minkowski space as well
and apply for this reason to a much wider class of observables
than the OPE underlying the QCD sum rules (\ref{qcd})
\footnote{There is a price to pay, however. Infrared renormalons if
applied directly in the Minkowski space do not allow
for model independent relations between power
corrections to different observables.
The reason is that the coupling $\alpha_s$
refers now to infrared region and is, therefore, of order unity.
As a result, all orders in $\alpha_s$ are equally important,
see (Zakharov, 1998). This is the same ``collapse''
of the perturbative expansion on the level of the power
corrections illustrated above on the example of the potential $V(r)$.}.
The link to QCD is again through the bald replacement of the
non-analytical in $\lambda^2$ terms by corresponding powers of $\La$:
\begin{eqnarray}
\alpha_s\sqrt{\lambda^2}&\rightarrow &c_1\La \nonumber\\
\alpha_s\lambda^2\ln\lambda^2&\rightarrow & c_2\La^2~\ldots
\end{eqnarray}
where $c_{1,2}$ are some coefficients treated as phenomenological
parameters.
The gluon condensate appears now merely as one of
the terms in this sequence.

The phenomenology based on such rules turns successful
(see, e.g., Webber (1999) and references therein).
The problem is not so much a lack of success but rather
too much of overlap (Akhoury and Zakharov, 1996)
with old-fashioned hadronization models, 
like the tube model. 

{\it To summarize:} there exist simple techniques which
allow to parameterize the infrared sensitive parts of the Feynman graphs
directly for observables in the Minkowski space. 
The success of this, most naive approach reveals that
at least in the cases when the technique applies the
nonperturbative effects reduce to a simple amplification
of infrared sensitive contributions to the Feynman
graphs.

\section*{Direct instantons}

This simple picture is not universally true, however.
First examples when it fails were
found as early as in 1981 (Novikov et.al.)
Namely, there exist channels where the infrared sensitive corrections
described
above cannot be the whole story.
To give counter-examples, one can either rely on the analysis
of experimental data or on theoretical methods which would allow
to evaluate the power corrections without using the OPE.
Both ways were exploited to demonstrate
that there exist corrections
which can be very
important numerically and which go beyond the picture described above.

In particular, in the $0^+$-gluonium channel one can 
establish a low energy theorem relating the value of
the correlation function at $Q^2=0$ to the same gluon
condensate $<\alpha_s(G_{\mu\nu}^a)>$.
Moreover, one can utilize this knowledge to
subtract the dispersion relations and convert
the $\Pi (0)$ into a power correction at large momenta.
As a result the following sum rules arise:
\beqn
&& \int ImG(s)\exp{(-s/M^2)}ds/s~\approx~
 \nonumber\\
&&\!\!\!\!\!\!\approx G(M^2)_{\rm parton ~model}
\left[1 +{\langle 0|\alpha_s(G^a_{\mu\nu})^2|0\rangle\over M^4}
\left(-{2\pi^2\over\alpha_s(M^2)}+{16\pi^2\over b_0\alpha_s^2(M^2)}
\right)\right]\!\!\label{stranger}
\eeqn
where
\be
G(Q^2)\equiv i\int\! d^4x\; e^{iqx} \;
\langle 0|\; T\{ (G^a_{\mu\nu}(x))^2,(G^a_{\mu\nu}(0))^2\}\; |0\rangle.
\ee
To reiterate, the power correction proportional to the gluon condensate
originates here from two sources. First, there is a standard OPE
correction (see (\ref{qcd})) and, second, the one evaluated
via a low energy theorem specific
for this particular channel. The correction which
is not caught by the standard OPE is about (20-30) times larger!

Thus, if we characterize the scale where the asymptotic freedom 
gets violated by the power correction by the value $M^2_{crit}$ 
as is mentioned above, then $M^2_{crit}$ differs 
drastically in some channels:
\be
(M^2_{crit})_{\rho-meson}~\approx~0.6~\mbox{GeV}^2,~~~
(M^2_{crit})_{0^+~gluonium}~\approx~15~\mbox{GeV}^2.\label{gluonium}
\ee
We see that the proof of the low-energy theorem brought a proof of existence
of qualitatively different scales in the hadron physics
(Novikov et.al., 1981).

One may call the $0^+$-gluonium channel exceptional
because of the appearance of this new correction.
In case of the $\pi$-meson channel, it was also possible to show that
the standard OPE corrections are not big enough to match
the contribution of the pion into the dispersive part of the sum rules.
Moreover, there arises a kind of hierarchy
of exceptional channels, or of $M_{crit}^2$.

This hierarchy could be explained qualitatively in terms
of transitions of the corresponding currents directly to instantons,
see Fig. 5 (Novikov et. al., (1981), Geshkenbein and Ioffe (1980)).
\begin{figure}[h]
  \begin{center}
    \leavevmode
    \epsfig{figure=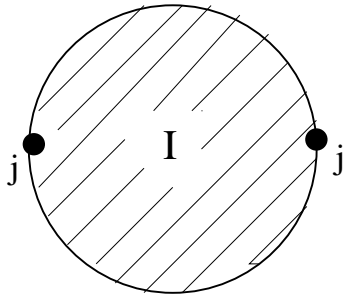,width=4.5cm}
    \caption{Direct instantons. One substitutes instanton fields, both
bosonic and fermionic
into the currents and integrates over the instanton sizes.}
    \label{fig:fig5}
  \end{center}
\end{figure}
At large $Q$ the direct instanton contribution dies off 
fast, like $(Q^2)^{-4.5}$. However, at intermediate $Q^2$
these corrections may become important first because
of a big overall
 coefficient.
Unfortunately, one cannot go far
with clarification of the situation by using only analytical means.
Nowadays one relies on the 
model of instanton liquid
which allows for a much more quantitative treatment
of the instanton effects
(for a review and further references see (Shuryak and Schafer, 1997).
The model appears to be successful phenomenologically.

\section{BEYOND THE OPE.}

\section*{Elusive effects of confinement}

Looking backward, it still remains a mystery
whether any specific confinement effects are revealed
through the power corrections
discussed so far. Indeed, consider the vacuum of pure
gluodynamics. It is known from lattice measurements
that external heavy quarks are confined by this medium.
On the other hand, the effects included into the sum rules so far
do not seem to encode the confinement.
Indeed, the perturbative QCD
resembles ordinary bremsstrahlung in QED.
The gluon condensate, as well as other newly found power corrections
can be detected by introducing a fictitious gluon mass which
is not sensitive to the non Abelian nature of gluons at all.
Finally, instantons are known not to ensure the confinement either
(Chen et.al., 1998).

In an attempts to find power corrections related more
directly to the physics of confinement one can turn to
the Abelian Higgs model (AHM)
which underlies the dual superconductor model
of the confinement (Mandelstam (1974), Nambu, (1974); 
Polyakov (1975),'t Hooft (1976)). 
The basic idea behind the model is that
the properties of the QCD vacuum are similar to the properties
of an ordinary superconductor. Indeed, if a pair of magnetic charges
is introduced into a superconductor, the potential energy
of the pair would grow linearly with the distance $r$ at large distances:
\be
\lim_{r\to \infty}{V(r)}~\approx~\sigma_{\infty}\cdot r\label{lingrowth}
\ee
where $\sigma_{\infty}$ is the tension of the Abrikosov-Nielsen-Olesen string.
In QCD, a similar phenomenon was postulated to happen,
with a change of magnetic charges to (color) electric, or dual charges.

If one introduces a pair of
external magnetic charges into the vacuum of the AHM model
in the Higgs phase
then the potential grows at large distances, see Eq. (\ref{lingrowth}).
The scale of distances is set up by the inverse masses
of the vector and scalar fields, $m^{-1}_{V,S}$. This growth
of the potential is due to the Abrikosov-Nielsen-Olesen
strings.

Consider now {\it short} distances, $r\ll m^{-1}_{V.S}$.
Then the Coulomb like interaction dominates. However, there
is a stringy correction to the potential at any small
distances (Gubarev et. al., 1998):
\be
\lim_{r\rightarrow 0}V(r)~=~-{Q_M^2\over 4\pi r} +\sigma_0r\label{sd}
\ee
where $Q_M$ is the magnetic charge. The ANO string is a bulky object on this
scale and is not responsible for the linear correction. Instead,
the stringy potential at short distances is due to an infinitely thin
topological string which connects the magnetic charges and which
is defined through vanishing of the scalar field along the string.
Thus, it turns out that at least in this model
the confined charges learn about confinement already at small distances
because of the short strings which are in fact
seeds for future
confining ANO strings. Amusingly enough,
it demonstrates that at short distances
a dimension two quantity is not necessarily the gluon mass squared
but could be a string tension as well.
Eq. (\ref{sd}) is the main result which we are going to substantiate
in this part of the lectures. However, we need to make a few steps
before we can explain (\ref{sd}). Also, we shall spend some time later
to discuss the validity of the OPE.

\section*{Field theoretical solenoid.}

A key element in the conjectured mechanism of the quark confinement
is the Abrikosov-Nielsen-Olesen string (Abrikosov. 1957), 
(Nielsen and Olesen 1973). 
Although eventually it will {\it not} be this string that
captures our attention, it is useful to begin with a brief
review of the ANO string.

The strings are classical solutions to the Abelian Higgs model (AHM).
The model describes  
a gauge field $A_{\mu}$ interacting with a charged scalar field $\Phi$
as well as self-interactions of the scalar field.
The corresponding action is:
\be\label{AHM_action}
S= \int d^4x \left\{
\frac{1}{4e^2} F^2_{\mu\nu} + \frac{1}{2} |(\diff - i A)\Phi|^2 + 
\frac{1}{4} \lambda (|\Phi|^2-\eta^2)^2
\right\}
\ee
where $e$ is the electric charge, $\lambda,\eta$ are constants and
$F_{\mu\nu}$ is the electromagnetic field-strength tensor,
$F_{\mu\nu}\equiv\diff_{\mu}A_{\nu}-\diff_{\nu}A_{\mu}$.
The scalar field 
condenses in the vacuum, $\ve =\eta$, and the physical 
vector and scalar particles are massive, $m^2_V=e^2\eta^2, m_H^2= 2
\lambda \eta^2$. In the perturbative regime, interaction of the massive
scalar and vector particles can be readily calculated order by order.

There exists, however, a topologically non-trivial stringy solution 
to the classical equations of motion which
possesses a cylindrical symmetry and is characterized by a
finite energy per unit of its length. In a way, it realizes an
analog of a solenoid in field theory.
The key element is to look for a solution with
a non-vanishing electric current
\be
j_{\mu}~=~e\left(\Phi^{*}(\partial_{\mu}\Phi)
-(\partial_{\mu}\Phi^{*})\Phi\right)
.\ee
Moreover, if one chooses the scalar field of the form,
\be
\Phi({\bf r})~=~e^{in\phi} f(\rho )\label{scalar}
\ee
then the electric current is circular and reminds, therefore, the
solenoid current. The value of $n$ in Eq. (\ref{scalar})
is integer for the field $\Phi$ to be a unique function
of the coordinates.
Since the current has only a nonvanishing $j_{\phi}$ component,
the functional form of the matching vector potential is
\be
{\bf A}({\bf r})~=~\hat{{\bf e}}_{\phi}{A(\rho)\over \rho}.\label{vector}
\ee
The central question is whether it is possible to ensure finiteness of the
energy (per unit length) with the ansatz (\ref{scalar}), (\ref{vector}).
The behavior of the fields at $\rho\to 0$ and $\rho \to \infty$ is
crucial at this point and it is easy to check that the
conditions
\beqn\label{asympt}
f(0)~=~A(0)~=~0~~~\\ \nonumber
f(\infty )~=~\eta~~~\\ 
A(\infty)~=~{n\over e} \nonumber 
\eeqn
suffice to make the energy finite.

From these conditions alone, one can derive the magnetic 
flux carried by the ANO string. Indeed,
\be
\int{\bf H}d{\bf S}~=~\oint A_{\mu}dx_{\mu}~=~{2\pi\over e}n
,\ee
where to evaluate the latter 
integral we used the asymptotic value of $A_{\phi}$, see Eq. (\ref{asympt}).
Thus, the flux is an integer of a minimal magnetic monopole charge.
It is obvious then that the strings can end up with monopoles.

Explicit form of the functions $f(\rho), A((\rho)$ introduced above
can be found by solving the classical equations of motion:
\beqn
 \bigtriangledown^2 A_{\mu}={ie\over 2}
\left(\Phi^{*}\partial_{\mu}\Phi-\Phi\partial_{\mu}\Phi^{*}\right)
-e^2A_{\mu}\\ \nonumber
 \bigtriangledown^2 \Phi+2ieA_{\mu}\partial^{\mu}\Phi-e^2A_{\mu}A^{\mu}
  =\lambda\eta^2\Phi-\lambda\Phi|\Phi|^2
\eeqn
In particular, one concludes from these equations that the function
$f(\rho)$ approaches its asymptotic value
at $\rho\to \infty$ exponentially, as $\exp(-m_H\rho)$.
The magnetic field falls off exponentially, as $\exp(~-m_V\rho)$.
Notice, however, that the vector potential $A_{\phi}$ falls off 
only as an inverse power of $\rho$.
There is no field strength tensor associated with
the asymptotic value of $A_{\phi}$.
As for the full
functions $f(\rho), A(\rho)$ they can be obtained by numerical methods.

Details of the solution depend on the ratio of the masses, $m_V$ and $m_H$.
Two limiting cases are of special interest. In the so called Bogomolny limit,
\be
m_V~=~m_H=~m, 
\ee
the equations simplify greatly and can be solved in fact analytically,
as a series in $m\rho$. 
Another useful case is the so called London limit,
\be
m_H^2~\gg~m_V^2
\ee
In this limit, the scalar field is frozen as $\Phi=\eta$ and one can neglect
in most cases the fluctuations of the field $\Phi$ around its vacuum value.
However, the string tension, or energy per unit length becomes
logarithmically divergent in this limit:
\be
\sigma_{\infty}~\approx~{\pi m_V^2\over 2e^2}\ln{m_H^2\over m_V^2} 
\label{tension}
\ee
Indeed, without the Higgs field one cannot construct a solution 
with a finite string tension. 

From the physical point of view,
the most important manifestation of the ANO string is 
the linearly growing potential energy for an external \ma pair
(\ref{lingrowth}),
when the distance $r$ is much larger than the inverse masses,
$m_{H,V}^{-1}$. Thus, the magnetic charges are confined.
This property of the Abelian Higgs model is the basis for
the dual superconductor mechanism of the quark confinement.
Namely, one assumes that the QCD vacuum is similar to the vacuum of
the AHM in the sense that a scalar field with non-vanishing
magnetic charge condenses in the QCD vacuum. 
Then the (color) electric charge is confined.

Note that once we allowed for point-like monopoles, we should have 
considered
the energy of the Dirac string as well. Let us discuss this issue on 
another example, that is compact $U(1)$ gauge theory.

\section*{Compact Photodynamics.}

In this section we will outline the paper
(Polyakov, 1975) which, from the point of view relevant to the present
lectures, demonstrates that there can exist a very different source of
nonperturbative effects than those we discussed so far.

The action of the theory we are going to consider is very simple:
\be
S~=~{1\over 4 e^2}\int d^4x F^2_{\mu\nu}\label{action}
\ee
where $F_{\mu\nu}$ is the abelian field strength tensor, 
$F_{\mu\nu}=\partial_{\mu}A_{\nu}-\partial_{\nu}A_{\mu}$.
The action (\ref{action}) is that of free photons
and at first sight nothing interesting can come out from
this theory. 
In particular, if we introduce external static
electric charges as a probe, their potential energy would be given by
one-photon exchange without any corrections.

However, we shall see in a moment that, in a particular formulation,
the theory admits also magnetic monopoles. Hence,
a few preliminary words on the monopoles.
Monopoles have magnetic field similar to the electric field of a charge:
\be
{\bf H}~=~q_M {{\bf r}\over 4\pi r^3}
\ee
Then the flux of the magnetic field through a surface surrounding
the monopole is:
\be
\Phi~=~q_M
\ee
On the other hand, because of the equation $div~{\bf H}=0$ 
the magnetic flux is conserved. 
Thus, the magnetic monopole cannot exist by itself
and one assumes that there is a string
which is connected to the monopole and which brings in the flux.
Moreover, to make the string invisible
one assumes that the string is infinitely thin. Finally, to
avoid the Bohm-Aharonov effect
one imposes the Dirac quantization condition,
\be
e\oint A_{\mu}dx_{\mu}~=~e\int {\bf H}\cdot d{\bf s}~=~2\pi\cdot n\;,
~~~\mbox{or}~~~
q_M~=~{2\pi\over e}n
\ee
Also, we ask for the energy (or action) associated with the  
Dirac string to vanish. Only then energies of the electric and 
magnetic charges 
are similar. We shall return to discuss the issue of the
energy of the Dirac string later. 

Now, the Dirac strings may end up with monopoles. The action
associated with the monopoles is not zero at all but rather
diverges in ultraviolet, since
$$ \int {d^3r\over 4\pi}{\bf H}^2~\sim~{1\over e^2a}$$
where $a$ is a (small) spatial cut off. If the length
of a closed monopole trajectory is of order L, then the
suppression of such configuration due to
a non-vanishing action is of order
\be
e^{-S}~\sim~\exp~(-const~L/e^2)
.\ee
On the other hand, there are different ways to organize a loop of
length $L$. This is 
the entropy factor. It is known to grow exponentially with $L$ as
$\sim \exp~(~const'L)$.

Thus, one comes to the conclusion 
(Polyakov, 1975) that at some
$e = e_{crit} \sim 1$
there is a phase 
transition corresponding to condensation of the monopole loops.
As a result, if external electric charges are introduced as a probe,
their potential energy grows with distance, $V(r)\sim r$ and
they are confined.
\footnote{This confinement mechanism and confinement mechanism of
magnetic monopoles in AHM discussed in the
previous section are interrelated. It can be shown in
the lattice regularization that compact $U(1)$ theory is dual to
AHM in which the bare Higgs and gauge boson mass are infinite.}

To complete the presentation we should explain how one should
understand the theory (\ref{action}) that it would imply
a vanishing action for the Dirac string.

The crucial point is to define the theory
by means of a lattice regularization.
Then the action can be understood as a sum over plaquette
actions:
\beqn
S~=\sum{1\over 2e^2}\left(1-~Re~ \exp({i}\oint A_{\mu}dx^{\mu})
\right)~=~
\nonumber \\
\sum{1\over 2 e^2}\left(1-Re~ \exp({i}
F_{\mu\nu}d\sigma_{\mu\nu})\right)~=~
\sum{1\over 2 e^2}\left(1-\cos(F_{\mu\nu}d\sigma_{\mu\nu})\right),
\eeqn
where the sum is taken over all plaquettes.
In the continuum limit one reproduces of course 
the action (\ref{action}). However, from the intermediate steps it is clear
that the action admits for a large jump in $F_{\mu\nu}$:
\be
F_{\mu\nu}~\rightarrow~F_{\mu\nu}+2\pi\delta (\sigma_{\mu\nu})\label{dstring}
\ee
where the $\delta$-function on the surface is defined
as $\delta(\sigma_{\mu\nu})d\sigma^{\mu\nu} =1$ (no summation over $\mu,\nu$).
The second term in Eq. (\ref{dstring}) exactly corresponds to the Dirac
string. Thus, the Dirac strings have no action in the compact version
of the $U(1)$ gauge theory. 

Note that the UV scale $1/a$
is the only scale at the model. Thus, it rather exists only as a lattice
theory. 
The confining potential sets in for $e > e_{crit}$ at all the distances
and in this sense the situation does not imitate QCD.
There is a possibility that a truly continuum theory can be
defined  near $e=e_{crit}$ but we would not discuss this topic here
(for a recent discussion and further references see, e.g., Jersak et.al
(1999))

\section*{Topology of gauge fixing}

The phenomenon of the monopole condensation is 
well established in QCD through numerical simulations on
the lattice, for review see (Chernodub et. al., 1998),
(Di Giacomo, 1998)..

The path to understanding magnetic monopoles in gluodynamics necessarily
goes through the paper by 't Hooft (1980)
where the monopoles were introduces as local objects and on general
topological grounds. It is convenient to start, however, with 
the Abelian Higgs model, see Eq. (\ref{AHM_action}), which contains 
rather strings than monopoles, to learn a new technique.
The gauge transformation rotates the phase of the 
charged field $\Phi$:
\be
\Phi~\rightarrow~e^{\i\phi}\Phi
.\ee
In particular one can rotate $\Phi$ in such a way that
\be
Im\Phi~=~0 \label{unitary}
\ee
which defines what is called the unitary gauge. The reason for the name 
is that
if $<\Phi>\neq 0$ then 
the excitation spectrum in this gauge consists of physical massive vector
and scalar particles, with no ghosts. 

The crucial point is that, as remarked by
't Hooft, the condition (\ref{unitary}) fixes the gauge uniquely unless
the real part of $\Phi$ is also vanishing.
Thus, there are exceptional points characterized by the equations
\beqn
Im~\Phi~=~0 \nonumber \\
Re~\Phi~=~0\label{strings}
.\eeqn
Viewed as two equations in four dimensional world, these conditions 
define a world sheet, or trajectory of a string.
Note that the definition (\ref{strings}) is local in the sense
that it defines an infinitely thin string. 

The definition does not give any
immediate clue as to whether
the strings (\ref{strings}) are important dynamically. 
However, one may notice that
$\Phi$ does vanish on the central axis of the ANO string, see (\ref{asympt}).
If one makes such an identification then it is easy to conclude that
the world sheets (\ref{strings}) are either closed, corresponding to closed
ANO strings, or end up with monopoles.

In case of a nonabelian theory, one introduces first
the so called $U(1)$ projection which is nothing else but (partial) 
gauge fixing. Namely, one chooses a field in an adjoint representation.
If we concentrate of $SU(2)$, then this field is a triplet $\Phi^a$.
The field $\Phi^a$ could be a component of the nonabelian field strength
tensor $F^a_{\mu\nu}$ or a composite operator. 
Then one chooses the gauge to rotate $\Phi =
\Phi^a \frac{\sigma^a}{2}$ to, say, the third
direction in the color space:
\be
\Phi~\rightarrow~ \Omega\Phi \Omega^{-1}~=~\Phi^3\frac{\sigma^3}{2}
\label{three}
\ee
where $\Omega$ is the matrix of the gauge transformation.
The condition (\ref{three}) fixes the gauge only up to
rotations around the third axis in the color space, that is up
to a remaining $U(1)$ symmetry.

What is crucial for introduction of the monopoles is that
the condition (\ref{three}) can be implemented everywhere except for
the points 
\be\Phi^a\equiv 0 \label{monopole}.\ee
Thus, the exceptional points are given 
by solutions to 
three equations in the four-dimensional space and specify 
therefore a trajectory.
Actually, this is a monopole trajectory. To prove this, one follows
exactly the same sequence of steps as in establishing the relation
of the 't Hooft-Polyakov monopole to the Dirac magnetic monopole,
see ('t Hooft, 1974). 
Indeed, 
in Georgi-Glashow model we diagonalize the higgs triplet $\Phi^a$
as in (\ref{three}) and $|\Phi|^2$  vanishes at the  center of a 
't~Hooft-Polyakov monopole. And the basic observation is
that the algebra associated with the rotation of the 
't~Hooft-Polyakov monopole to the Dirac monopole
is in fact uniquely determined by this circumstance. A detailed derivation
can be found, e.g., in (Simonov, 1996) or in (Chernodub et.al., 1998).
 
Note that the definition of the monopole (\ref{monopole})
is pure topological, the same as the definition of the string above,
see (\ref{strings}).
It does not at all imply existence of a classical monopole solution. 
However, it provides with a local definition of a magnetically charged
field $\Phi_M(x)$ and allows to study the monopole condensation.
Note also that the definition of the monopoles depend on the choice of
the auxiliary field (operator) $\Phi^a$.
The dual superconductor model of confinement is confirmed numerically 
most impressively within the so called maximal Abelian gauge.
In this case, one fixes gauge, up to a $U(1)$ subgroup
by minimizing the value of $(A_{\mu}^1)^2+(A^2_{\mu})^2$, where 
$A_{\mu}^{1,2}$ are 
components of the vector potential in the color space. The 
remaining $U(1)$ symmetry is the phase rotation of the complex field
$A^1_{\mu}+iA^2_{\mu}$. We note all this only in passing, and
the reader is advised to consult (Chernodub et. al., 1998) for any
detail.

{\it To summarize}: both strings in the Abelian Higgs model and monopoles
in gluodynamics can be defined locally in topological terms.
These definitions do not provide however, with any straightforward
way to evaluate the dynamical significance of these objects. 

\section*{Ultraviolet regularization and non-perturbative effects.}

Now we would like to introduce a new idea 
on the connection of the topological
strings and monopoles discussed in the preceding section
and nonperturbative {\it ultraviolet} divergences. 
Indeed, intuitively it is appealing to guess
that, if the string- or monopole-like 
solutions with finite energy exist only for theories with
scalar fields, similar point-like, or topological excitations build 
up on the gauge fields
alone have infinite action, generally speaking. Only if we accept
a prescription of an ultraviolet regularization which eliminates 
these divergences, such excitations may play a dynamical role. 
Although intuitively the idea looks appealing, we shall not be able to prove
it in its generality but rather illustrate it on examples.

Let us begin with the Dirac string. Naively, the energy $\epsilon_D$
is equal to:
\be
\epsilon_D~\sim~\int{\bf H}^2d^3r~\sim~{(Flux)^2\over A^2}A\int dl
,\ee
where $A$ is the cross section of the Dirac string.
Thus, $\epsilon_D$ diverges as $1/A$ (quadratically) in ultraviolet
once $A\to 0$.

Instead, we assumed $\epsilon_D=0$. There are two ways to justify
this assumption:

({\it i}) Lattice regularization, as discussed in the preceding subsection.

({\it ii}) Duality between electric and magnetic charges. Since the electric
charges do not have strings attached, the action associated with the 
Dirac string is postulated to vanish.

Note that while in case of the compact $U(1)$ we relied on
point ({\it i}),
while 
in case of the Abelian Higgs model we had to rely on ({\it ii}).

Turning to $SU(2)$, there is of course a possibility to introduce
a Dirac string associated with any $U(1)$ subgroup.
In this connection, let us mention another language for the
strings, that is singular gauge transformations,
for details see, e.g., Chernodub (1998). 
Gauge transformations can be specified in terms of a matrix $\Omega$:
\be
\Omega(x)~=~\exp\left(i\alpha^a(x){\sigma^a\over 2}\right)
\ee
where $\alpha^a (a=1,2,3)$ are parameters of the transformation
and $\sigma^a$ are the Pauli matrices.
Then
\be 
\hat{A}_{\mu}~\rightarrow~\Omega^{\dagger}(x)
\left(\hat{A}_{\mu}(x)- {i\over g}\partial_{\mu}\right)\Omega(x) 
 \ee
where $\hat{A}_{\mu}=A_{\mu}^a\sigma^a/2$.
Moreover for the field strength one has
\be
\hat{F}_{\mu\nu}~\rightarrow~\Omega^{\dagger}\hat{F}_{\mu\nu}\Omega
-{i\over g} \Omega^{\dagger}(\partial_{\mu}\partial_{\nu}-\partial_{\nu}\partial_{\mu})
\Omega
\ee
where we reserved for the possibility that the 
derivatives $\partial_{\mu,\nu}$ do not commute when applied to $\Omega(x)$.
The noncommutativity is in fact the definition of a singular 
gauge transformation.

Consider furthermore
\be
\Omega(x)~=~\left(\matrix{\cos{\gamma\over 2}&\sin{\gamma\over 2}e^{-i\alpha}\cr
-\sin{\gamma\over 2}e^{i\alpha}& \cos{\gamma\over 2}\cr}\right)
\ee
where $\alpha$ and $\gamma$ are azimuthal and polar angles, respectively.
Then it is straightforward to check that
\be 
\hat{F}_{\mu\nu}(\hat{A}^{\Omega})~=~
\Omega^{\dagger}\hat{F}_{\mu\nu}(\hat{A})\Omega- \;
\sigma^3 \;
{2\pi\over g} \;  
(\delta_{\mu,1}\delta_{\nu,2}-\delta_{\mu,2}\delta_{\nu,1})
\; \delta(x_1)\delta(x_2) \; \Theta(-x_3)\label{abelianstring}
\ee
In other words, we generated a Dirac string directed along the $x_3$-axis
ending at $x_3=0$ and carrying the color index $a=3$.

It is quite obvious that such Abelian-like strings are allowed by
the lattice regularization of the theory. However, we cannot get too
far using only Abelian-like field configurations. Indeed, imagine
that the string (\ref{abelianstring}) would go over into the same
Abelian monopoles at its end points. Then we would have the same 
estimates as in the case of the compact $U(1)$ but our
coupling now tends to zero in the ultraviolet,
$\lim_{a\to 0} g^2(a)=0$ and these field configurations are
strongly suppressed. 

The best qualitative picture for QCD based on the analogy with the compact
$U(1)$
seems to be as follows. Begin with a small lattice size $a$ but then
go to a coarser lattice, with corresponding renormalization of the coupling
a la Wilson.
Then once the effective coupling governing a $U(1)$ subgroup
reaches the value of $e^2_{crit}$ then one may think that the monopoles
are condensed, in the same way as in the $U(1)$ case.
There is a  semi-quantitative 
prediction based on this picture. Namely, it is natural
to assume that the running of the coupling is changed drastically at this 
point as far as further advance to the infrared is concerned.
Thus, we can speculate that the coupling is frozen at
the value
\be
g^2_{frozen}~\sim~2e^2_{crit}~\sim~2
\ee
which is not unreasonable phenomenologically\footnote{Note that the
lattice calculations show (Kato 1998, Chernodub 1999) that the
abelian monopoles of finite physical size
(``extended'' monopoles (Ivanenko 1990))
are important for confinement.}.

{\it To summarize}, we touched upon a few issues in this subsection.
First of all, we emphasized that introduction of the Dirac string 
implies, at least naively, a new quadratic {\it ultraviolet} divergence. 
We also argued that the Abelian-like
monopoles cannot be important in the non-Abelian case since the bare
coupling goes to zero. 
In a way, the non-Abelian monopoles are to be so to say empty at short
distances in the sense that the action is determined by contribution
of distances of order ${\La}^{-1}$ where the coupling is not small.
How this might happen, we shall explain in the next section.

\section*{Magnetic monopoles in the limit $m_H\to\infty$.}

Magnetic monopoles of finite energy are found 
('t Hooft (1974), Polyakov (1974))
as classical solutions 
to nonabelian field theories which include scalar fields.
Thus, it is a common question, how one can understand monopoles
in QCD when there are no scalar fields. We will provide a partial
answer to this question, arguing that the scalar fields are
replaced in some sense by singular gauge transformations
as far as short distances are concerned.

Begin with the Georgi-Glashow model which includes a tripled
$\Phi^a$ of scalar fields:
\be
L~=~{1\over 4}(F^a_{\mu\nu})^2+{1\over 2}(D_{\mu}\Phi^a)^2+
{1\over 8}\lambda((\Phi^a)^2-v^2)^2
\ee
Here,
\beqn
F^a_{\mu\nu}~=~\partial_{\mu}A_{\nu}^a-
\partial_{\nu}A_{\mu}^a+g\epsilon^{abc}A_{\mu}^bA_{\nu}^c\\
D_{\mu}\Phi^a~=~\partial_{\mu}\Phi^a+g\epsilon^{abc}A_{\mu}^b\Phi^c,
\eeqn
and $\lambda, v$ are constants.

Furthermore, one introduces functions $A(r)$, $\Phi(r)$:
\be
A_{\mu}^a~=~\epsilon^{\mu ab}r^b A(r),~~~\Phi^a~=~r^a\Phi(r).
\ee
The monopole solution is characterized by its asymptotic at $r\to \infty$:
\be
\lim{A(r)}_{r\to\infty}~=~-r^{-2}g^{-1},~~
\lim{\Phi(r)}_{r\to\infty}~=~v\cdot r^{-1}.\label{inft}
\ee
The mass of the monopole is:
\be
M_{monopole}~=~{4\pi v\over g}C(\lambda/g^2)\label{mm}
,\ee
where $C(\lambda/g^2)$ is in fact a very slow function of its argument,
as emphasized already by 't Hooft (1974). Eq. (\ref{mm}) implies then that
the mass of the monopole corresponds to the mass of a Dirac monopole
with an ultraviolet cut off at distances of order $r\sim m_V^{-1}$.

What is a crucial point for us now, is that the mass of the monopole
remains finite in the limit $\lambda\to \infty$ which is the same as 
the limit of infinite Higgs mass, $m_H^2=\lambda v$.
Thus, we may take the scalar mass as an ultraviolet cut off and hope to remove
the scalar from the spectrum altogether. Now we are dealing with QCD still having
a finite mass of monopole. Let us look closer (in the limit $m_H \gg m_V$)
why we do need the scalar field.

Consider distances $m_V^{-1}\gg r \gg m_H^{-1}$. The point
is that at such distances the field of the monopole can be nothing else
but nearly a pure gauge,
\be
F_{\mu\nu}^a~\approx~0\label{trivial}
.\ee
Indeed, only in this case the mass of the monopole would be insensitive 
to distances much smaller than $m_V^{-1}$, as testified by (\ref{mm}).
To find the solution more explicitly, let us look closer at the equations
of motion. Since we consider $r\gg m_H^{-1}$ we choose $\Phi(r)$
in its asymptotical form, $r\to \infty$, see (\ref{inft}).
Moreover, we make an ansatz:
\be
A(r)~\approx~{c\over g r^2}\label{firstterm}
\ee
where $c$ is a constant.

Then the equations of motion reduce to a simple algebraic equation
(see, e.g., 't Hooft (1974)):
\be
0~=~4cg^{-1}+6g^{-1}c^2+2g^{-1}c^3+2v^2r^2g(1+c).\label{mequation}
\ee
At the distances we are considering the product $gvr\ll 1$ and let us 
neglect for the moment the last term in (\ref{mequation}).
Then we have three solutions for the constant $c$:
\be 
c_1~=~-1,~~c_2~=~-2,~~ c_3~=~0.
\ee
The first solution is singled out by the fact that it
nullifies also the $r^2$ terms in (\ref{mequation}).
This is crucial at large $r$ and that is why $c_1$ is valid
asymptotically, see (\ref{inft}). 
However, this solution implies $(F_{\mu\nu}^a)^2\sim r^{-4}$ and a 
large contribution to the mass from distances $r\ll m_V^{-1}$ 
which is not allowed, see (\ref{mm}). 
On the other hand, if we choose $c_2=-2$ then Eq. (\ref{trivial})
is fulfilled and the contribution to the action from the non-Abelian
field vanishes.
Thus, at small distances we should turn rather
to the solutions $c_{2,3}$. To remove the contradiction
with the Eq. (\ref{mequation}) we 
include the next term in the expansion (\ref{firstterm})
\be
A(r)~\approx~{c\over g r^2}+dv^2\ln r,
\ee
where $d$ is a constant and the second term is a small 
corrections at the distances we are considering.
The constant $d$ can be readily found: 
$d=~ \pm \; g/3$.

Since $F_{\mu\nu}^a=0$, the potential
\be
A_{\mu}^a~=~-{2\over r^2}\epsilon^{\mu ab}r^b
\ee
is obtainable from $A_{\mu}^a\equiv 0$ by a gauge transformation:
\be
-{2\over r^2}\epsilon^{\mu ab}r^b{\sigma^a\over 2}~=~i (\Omega^0)^{\dagger}\partial_{\mu}
\Omega^0\label{ss}
.\ee
It is not difficult to find out the explicit form of the matrix $\Omega^0$:
\be
\Omega^0~=~i\vec{\sigma}\cdot\vec{n}\label{sing}
\ee
where $\vec{n}$ is the unit vector from the center 
of the monopole to the observation point. 

Now, it is clear that if we would introduce gauge rotations
(\ref{sing}) ad hoc, without scalar fields,
then we would conclude that the corresponding $A_{\mu}$
has a $\delta$-function singularity at the origin so that
not only $F_{\mu\nu}^a$ but the total mass  
$\sim\int (F_{\mu\nu}^a)^2d^3r$
is UV divergent. In this sense the only function of the scalar field is to 
smoothen the fields at the origin so that the contribution of the
singularity is in fact normalized to zero.
Now, the observation is that if we use the lattice regularization
and ask the question whether the singular (at $r=0$) potentials (\ref{ss})
are allowed or not, the answer would be in positive. Namely, the
center
of the singular monopole would fall into a center of a lattice cube
and this would imply that the singularity does not contribute
to the action in fact.

\section*{Short Strings.}

We proceed now to derivation
of the stringy correction (\ref{sd}) to the potential $V(r)$
within the Abelian Higgs model (Gubarev et. al., 1998).
Thus, we consider the following problem. Two external magnetic charges
are put at a distance $r$ in the vacuum of the Abelian Higgs model.
In particular, the vacuum expectation value of the scalar field $\Phi$
is non-vanishing. The problem is to find the potential energy $V(r)$
at distances much smaller than $m_H^{-1}, m_V^{-1}$.

The energy is determined in the classical approximation and,
at first sight, the problem is not much of a challenge.
The crucial point is that one has to impose an extra 
boundary condition, namely vanishing of the scalar field
along a mathematically thin line connecting the magnetic charges.
We have already mentioned this condition while
discussing
the topological strings (see Eq. (\ref{strings} and the related discussion).

The language of the Dirac string can be useful to
substantiate the point.
Indeed, the infinitely thin line discussed above is nothing
else but the Dirac string connecting the monopoles. The possibility of
its dynamical manifestations arises from the fact that the Dirac
string cannot coexist with $\Phi\neq 0$ and $\Phi$ vanishes along the
string. Indeed, self-energy of the Dirac string is normalized to
be zero in the perturbative vacuum. To justify this one can invoke
duality and ask for equality of self-energies of electric and magnetic
charges. Since the electric charge has no string attached, the
requirement would imply vanishing energy for the Dirac string.
However, if the Dirac string would be embedded into a vacuum with
$\ve\neq 0$ then its energy would jump to infinity since there is the
term $1/2|\Phi|^2A_{\mu}^2$ in the action and $A_{\mu}^2\rightarrow
\infty$ for a Dirac string.  Hence, $\Phi=0$ along the string and it
is just the condition mentioned above. In other words, Dirac
strings always rest on the perturbative vacuum which is defined as the
vacuum state obeying the duality principle. Therefore, even in the
limit $r\to 0$ there is a deep well in the profile of the Higgs field
$\Phi$. This might cost energy which is linear with $r$ even at small
$r$.

The next question is whether this mathematically thin
string realizes as a short physical string. Where by the physical
string we understand a stringy piece in the potential, $\sigma_0r$
at small $r$.
In other words, we are going to see whether the
stringy boundary condition implies a stringy potential.
To get the answer we solve classical equations
of motion. Let us note that the scheme of the calculation and
some numerical results for the
potential $V(r)$ can be found in a number of papers,
see, e.g., (Ball and Caticha, 1988). 
However, prior to (Gubarev et. al., 1998)
there were no measurements
dedicated specifically to small corrections to the Coulombic potential
at $r\rightarrow 0$.
  
We will consider the unitary gauge, $Im\Phi=0$.
Then
the most general ansatz 
for the fields consistent with the symmetries of the problem is:
\bea{ll}
\Phi=\eta f(\rho,z) &
Im f =0 \\
\\
A_{a}=\varepsilon_{a b} \hat{x}_b A(\rho,z) &
A_{0}=A_{3}=0 \\
\\
\rho=[x_a x_a ]^{1/2} & z=x_3 \qquad \hat{x}_a=x_a/\rho 
\qquad a=1,2 \ .
\eea

In the limit $r \to 0$ the Coulombic contribution becomes singular. 
The easiest way to separate the singular piece is to change the
variables $A=A_d+a$, where $A_d$ is the solution
in the absence of the Higgs field:
\be\label{dipole_potential}
A_d=\frac{1}{2\rho}\left[
\frac{z_{-}}{r_{-}} - \frac{z_{+}}{r_{+}}
\right],~~~
z_{\pm}=z \pm r/2 \qquad r_{\pm}=\left[ \rho^2 + z_{\pm}^2 \right]^{1/2}.
\label{change}\ee
Let us introduce also a new variable
$\kappa=\sqrt{2\lambda}/e =  m_H/ m_V$ and measure all dimensional
quantities in terms of $m_V = e\eta$. 
Upon the change of variables (\ref{change}) the energy functional and
the classical equation of motion take the form:
\bea{l}\label{energy-2}
E(r)=E_{self}-\pi/r + \tilde{E}(r),
\\
\\
\tilde{E}(r)=\frac{\pi}{e^2}
\int\limits^{+\infty}_{-\infty}dz \int\limits^{+\infty}_{0} \rho d\rho
\left\{
[\frac{1}{\rho}\diff_\rho (\rho a)]^2 + [\diff_za]^2 +
\right.
\\
\left.\rule{0.3\textwidth}{0.0mm}
+[\diff_\rho f]^2 + [\diff_z f]^2 + f^2(a+A_d)^2 + \frac{1}{4}\kappa^2(f^2-1)^2
\right\}
\eea

\beq\label{eq-motion-2a}
\diff_\rho \left[ \frac{1}{\rho} \diff_\rho (\rho a)
\right] + \diff^2_z a = f^2 (a+A_d)
\eeq

\beq\label{eq-motion-2b}
\frac{1}{\rho}\diff_\rho \left[ \rho \diff_\rho f \right] + \diff^2_z f = f (a+A_d)^2 
+ \frac{1}{2}\kappa^2 f (f^2-1)
\eeq

The energy functional has been minimized numerically.
The numerical results (Gubarev et. al., 1998)
for various
$\kappa$ values clearly demonstrate that there is a linear
piece in the potential even in the limit $r\ll 1$.
The slope $\sigma_0$ at $r\to 0$ was defined by the fitting 
the numerical data to:
\beq\label{fitting-function}
\tilde{E}_{fit}(r)= C_0 \left(\frac{\displaystyle 1-e^{-r}}{r}-1\right) + 
(\sigma_0 + \frac{1}{2}C_0)r =
\sigma_0 r + O(r^2),
\eeq
The resulting slope $\sigma_0$ depends smoothly on the value of $\kappa$,
For the purpose of orientation let us note that for $\kappa=1$
the slope of the potential at $r\to 0$ is the same
as at $r\to \infty$ That is, within error bars:
\be
\sigma_0~\approx~\sigma_{\infty}\label{simple}
\ee
where $\sigma_{\infty}$ determines the value of the potential at large $r$.

To summarize, existence of short strings has been proven 
in the classical approximation to
the Abelian Higgs model. 
The linear piece in the potential at small distances 
reflects the boundary condition that $\Phi=0$
along the straight line connecting the
monopoles.

\section*{Strings vs OPE.}

Let us emphasize that both in case of the compact $U(1)$ and in case of
the Abelian Higgs model the OPE does not apply to study the power
corrections. 

In case of the compact $U(1)$ it is especially evident that
the OPE does not work. Indeed, as we emphasized many times,
the use of the OPE allows to parameterize infrared sensitive parts of
the Feynman graphs. But there are no Feynman graphs at all now
because
we are considering the Lagrangian of a free gauge field. 
Nevertheless, the answer depends on the value of the coupling $e^2$.
If $e^2$ is higher than the critical value
$e^2_{crit}\sim 1$ then the system is in confinement phase (see above).
As a result the physics is changed at all the distances. Indeed,
the ultraviolet scale $a$ (which may be thought of as the lattice size)
is the only scale in the problem
since the coupling does not run. As a result, the potential
between external electric charges becomes linear at all the distances.
One photon exchange is not seen at all. Clearly, this observation
is equivalent to saying that the OPE does not work at any distance
because the OPE is just the assumption that at short distances one sees
particle exchange. Thus, we may say that the example of the compact
$U(1)$ is too strong. If $e^2>e^2_{crit}$ then there is no
particle exchange at all. While in QCD we believe that
perturbation theory applies to describe the leading effects at large 
$Q^2$.

Turn now to the AHM where
we can check OPE against the direct calculation of the potential
in the preceding subsection.
At first sight, existence of the $1/Q^2$ corrections in case of the AHM
is not surprising since now there is an operator of dimension $d=2$,
that is $|\Phi|^2$. It is worth emphasizing therefore that a bit more
careful analysis demonstrates that the power correction 
to the potential which we found in the AHM also contradicts the OPE.

Moreover, it is the existence of short strings that is manifested
also through breaking of the standard operator expansion. Indeed, above we
found the potential in the classical approximation. In this
approximation the potential is usually directly related to the
propagator $D_{\mu\nu}(q^2)$ in the momentum space,
\be V(r)=\int {d^3r\over (2\pi)^3} \; e^{i{\bf q\cdot r}}\;D_{00}({\bf q}^2)
\label{aa}\ee
Moreover, as far as $q^2$ is in
the Euclidean region and much larger than the mass parameters,
the propagator $D_{\mu\nu}(q^2)$ 
can be evaluated by using the OPE.
Restriction to the classical approximation implies that loop contributions
are not included. However, vacuum fields which are soft on the
scale of ${\bf q}^2$ can be consistently accounted for in this way
(for a review see, e.g., Novikov et.al. (1985)).
This standard logic can be illustrated by an
example of the photon propagator connecting two electric currents.
Modulus longitudinal terms, we have: 
\be \label{prop}
D_{\mu\nu}(q^2) ~=~\delta_{\mu\nu}
\left({1\over q^2}+{1\over q^2}e^2\langle \Phi^2\rangle {
1\over q^2}+{1\over q^2}e^2\langle \Phi^2\rangle {
1\over q^2}e^2\langle \Phi^2\rangle {
1\over q^2}+...\right)~=~{\delta_{\mu\nu}\over q^2-m_V^2}.       
\ee
Thus, one uses first the general OPE assuming $|q^2|\gg e^2\Phi^2$
then substitutes the vacuum expectation of the Higgs field 
$\Phi$ and upon summation of the whole series of the power corrections
reproduces
the propagator of a massive particle. 
The latter can also be obtained 
by solving directly the classical equations of motion.

This approach fails, however, 
if there are both magnetic and electric charges 
present. In this case, one can choose the Zwanziger formalism
(Zwanziger, (1971), Brandt (1978)) 
and work out an expression for propagation of a photon 
coupled to magnetic currents following 
literally the same steps as in (\ref{prop}).
In the gauge $n_{\mu}\tilde{D}_{\mu\nu}=0$ the result is
well known:
\be \label{wrong}
\tilde{D}_{\mu\nu}(q,n)~=~{1\over q^2-m_V^2}\left(\delta_{\mu\nu}-
{1\over (qn)}(q_{\mu}n_{\nu}+q_{\nu}n_{\mu})+{q_{\mu}q_{\nu}\over(qn)^2}+
{m_V^2\over (qn)^2}(\delta_{\mu\nu}n^2-n_{\mu}n_{\nu})\right)   
.\ee
Here the vector $n_{\mu}$ is directed along
the Dirac strings attached to the magnetic charges 
and there are general arguments
that there should be no dependence of physical
effects on $n_{\mu}$. On the other hand, 
if the potential energy is given by the Fourier transform of (\ref{wrong})
then its dependence on $n_{\mu}$ is explicit.

Note that Eq. (\ref{wrong}) immediately implies that the standard OPE
does not work any longer on the level of $q^{-2}$ corrections.
Indeed, choosing $q^2$ large and negative does not guarantee now that
the $m_V^2$ correction is small since the factor $(qn)^2$ in the
denominator may become zero. Of course, appearance of the poles in 
$(qn)$ in the longitudinal terms is not dangerous since they
drop due to the current conservation. However, the term proportional
to $m_V^2$ in (\ref{wrong}) cannot be disregarded and contribute, 
in particular, to (\ref{aa}).

The reason for the breaking of the
standard OPE is that even at short distances the dynamics of the short
strings should be accounted for explicitly. In particular, in the
classical approximation the string lies along the straight line
connecting the magnetic charges and affects the solution through the
corresponding boundary condition, see above. More generally, the OPE
allows to account for effects of vacuum fields, in our case for $\ve
\neq 0$. The OPE is valid therefore as far as the probe particles do
not change the vacuum fields drastically and the unperturbed vacuum
fields are a reasonable zero-order approximation.  In our case,
however, the Higgs field is brought down to zero along the string and
this is a nonperturbative effect. Thus, the stringy piece in the
potential $V(r)$ at $r\to 0$ is a nonperturbative correction which is
associated with short distances and emerges already on the classical
level.

\section{REVISITING THE PHENOMENOLOGY.}

\section*{Revisiting the quark potential at short distances.}

We have shown above that in the AHM the quark potential 
contains a linear correction at short distances.
Moreover, the reason for the breaking of the OPE seems
to be a unified description of magnetic and electric charges.
Since the monopole condensation is established within the
abelian projection of QCD,
we would speculate that a similar picture for the potential
is true in QCD as well. Then we expect a linear piece
dominate over other non-perturbative corrections at short distances
and this would make it unnecessary to account for the
retardation effects discussed in part I. 

Because the theoretical foundation for the new picture is
not absolutely solid, phenomenological evidence would be
especially important at this point. 

Without going into detail, because of the space consideration,
let us mention that there are a few areas where measurements, especially
on the lattice, could clarify the situation:

({\it i}) Direct measurements of the potential $V(r)$
bring so far nonvanishing results for 
the effective string tension at short distances 
(Bali et al, 1995, Bali, 1999):
\be
\sigma _{0}~=~(1\div 6)\sigma_{\infty}
\ee

({\it ii}) There exist lattice
measurements of fine splitting of the $Q\bar{Q}$
levels as function of the heavy quark mass~\footnote{
We are thankful to A. Leonidov for bringing this issue to our attention
and for introducing to the literature.}.
The Voloshin-Leutwyler picture results in a particular
pattern of the mass dependence
of this splitting. Moreover, these predictions
are very different from the predictions based
on adding a linear potential to the Coulomb potential
at short distances. Numerically, predictions based on short strings with
$\sigma_0\approx\sigma_{\infty}$ (see above) are close
to the predictions obtained by Buchmuller and Tye (1980) within
a phenomenological model for the potential.

The results of most advances measurements of this type are
published in (Fingberg, 1998) and favor strongly the linear 
correction to the potential at short distances.

({\it iii}) There is an interesting evidence that the
nonperturbative fluctuations on the lattice responsible for the
confinement can be identified as the so called P vortices 
(Faber et. al., 1999).
If one measures the quark potential due only to these vortices, the
numerical results seem  to indicate that the slope of the potential is 
the same at large and small distances, see (L. Del Debbio et. al., 1997).

({\it iv}) Analytical studies of the Bethe-Salpeter equation and comparison
of the results with the data about the charmonium spectrum
favor a nonvanishing linear correction to the potential
at short distances (Badalian and Morgunov, 1999).

Thus, it appears that existing experimental data favor a linear correction
to the heavy quark potential at short distances but much more work is needed
to finalize the results.

{\it To summarize}: the stringy correction to the Coulombic potential 
at short distances found in the Abelian Higgs model violates the 
operator product expansion. The search for a linear piece in the potential
at short distances in QCD is of great importance.
We listed above some preliminary indications that such a correction is
indeed present.

\section*{Revisiting correlation functions: tachyonic gluon mass at short distances.}

 The $1/Q^2$ corrections discussed in the previous section go beyond
the standard OPE. Detection of the new type corrections
through phenomenological studies 
would be of great interest. In this section we will discuss
phenomenology in terms of a tachyonic gluon mass which is assumed to mimic
the short-distance nonperturbative effects 
(Chetyrkin et al., 1998).

First, let us note that not all the $1/Q^2$ corrections in QCD are
associated with short distances. For example, in case of 
DIS the $1/Q^2$ corrections are coming from the 
IR region and perfectly consistent with the OPE.
Thus, the class of theoretical objects for which an observation
of the $1/Q^2$ corrections would signify going beyond the OPE is
limited. One example was the potential $V(r)$ discussed above.
Other examples are the correlator functions (\ref{correlator}) where
the IR-sensitive power corrections start with $Q^{-4}$ terms,
as is explained in detail in the Part I of the lectures. 

Thus, we concentrate on this set of variables and  
an interesting question is whether
there is room for introduction of sizable non-standard $1/Q^2$ 
corrections to the correlator functions. 
The answer seems to be in positive.

Even if one accepts that the non-standard $1/Q^2$ correction
to the
potential has already been observed (see the preceding section), 
it is far from being trivial
to evaluate the 
$1/Q^2$ corrections to other quantities.  
Qualitatively, however, one may hope that
introduction of a tachyonic gluon mass at short distances  
would imitate the effect of the 
$\La^2/Q^2$ corrections. Indeed, the linear term 
$\sigma_0\cdot r$ in the potential
at short distances
can be imitated by the Yukawa 
potential with a gluon mass $\lambda $:
\be
{4\alpha_s\over 6}\lambda^2~\sim~-~\sigma_0\label{mass}
.\ee
On the Born level, Eq. (\ref{mass}) is an identity.
However,
a tachyonic gluon mass can be consistently used at one-loop
level as well. 
Of course Eq. (\ref{mass}) may serve only for a rough estimate.

Looking for other applications, let us remind the reader, that
one of the basic quantities to be determined from the theory is 
the scale $M^2_{crit}$ at which
the parton model for the correlators
gets violated considerably via the power corrections. 
Now, a new term proportional to $\lambda^2$ is added
to the theoretical side of $\Pi_j(M^2)$ which becomes: 
\be
\Pi(M^2)\approx (parton~model)(1+{a_j\over \ln M^2}+{b_j\over M^4}+{c_j\over M^2})
\ee
where $c_j$ is calculable in terms of $\lambda^2$ 
(Chetyrkin et.al., 1998)\footnote{Further discussion of the 
short-distance tachyonic
gluon mass can be found in (Zakharov, 1998), (Simonov, 1999), 
(Huber et.al., 1999).}:
\be
c_{\pi}\approx 4c_{\rho}={4\alpha_s\over 3\pi}c_{gluonium}=
{4\alpha_s\over \pi}\lambda^2.
\ee
Phenomenologically,
in the $\rho$-channel there are severe restrictions (Narison, 1992)
on the new term $c_{\rho}/M^2$:
\be
c_{\rho}~\approx~-~(0.03\div 0.07)~\mbox{GeV}^2\label{constr}
.\ee
Remarkably enough, the sign of $c_{\rho}$
does correspond to a tachyonic gluon mass
(if we interpret $c_{\rho}$ this way).
Moreover, when interpreted in terms of $\lambda^2$ the constraint (\ref
{constr}) does allow for a large $\lambda^2$, say, $\lambda^2=-0.5~\mbox{GeV}^2$.

As for the $\pi$-channel one finds now a new value of $M^2_{crit}$
associated with $\lambda^2\neq 0$:
\be
M^2_{crit}(\pi-channel)\approx~4\cdot M^2_{crit}(\rho-channel).\ee
It is amusing that just this value of $M^2_{crit}$ was
found in (Novikov et.al., 1981)
from an analysis of the pion contribution
and it cannot be explained within the standard QCD sum rules.
Moreover, the sign of the correction in the $\pi$-channel 
is what is needed for phenomenology (Novikov et.al., 1981).
Fixing the value of $c_{\pi}$ to bring the theoretical $\Pi_{\pi}(M^2)$
into agreement with the phenomenological input one gets
\be
\lambda^2~\approx~-0.5~ \mbox{GeV}^2
.\ee
Finally, with this value of $\lambda^2$ in hand,
we can determine the new value of $M_{crit}^2$ in the 
scalar-gluonium channel and, again, it turns to be what is 
needed for the phenomenology,
see Eq. (\ref{gluonium}).

Further checks could be provided by measurements of various correlation
functions on the lattice, similar to the measurements
reported in (Chu et. al., 1993). It would be most interesting 
to try to disentangle the $1/Q^2$ corrections discussed here from
the effects of the direct instantons mentioned in Part 1.
Let us give two particular examples:

({\it i}) In case of instantons, the correlation function of
the tensor currents, which essentially coincide with the
energy-momentum tensor are protected against large corrections.
Indeed, it is well known that the energy-momentum tensor vanishes
on the instantons. If the tachyonic mass is responsible for large corrections,
the tensor gluonic current is not singled out in any way.

({\it ii}) It would be very important to measure the sum
$$\Pi_{\sigma}(x)+\Pi_{\pi}(x)$$
at critical values of $x$ where the violation of the
asymptotical freedom becomes noticeable.
The point is that the direct instantons drop off from 
this sum to the first order
while the effect of the tachyonic mass would add up. The existing data
(Chu et. al., 1993),
taken literally leave ample space for the tachyonic mass. Measurements
with improved accuracy are highly desirable.

It is worth emphasizing also that the $\lambda^2$ terms represent 
nonperturbative physics and limit in this sense the range of applicability
of pure perturbative calculations. 
This nonperturbative piece may well be 
much larger than some of the perturbative corrections which are 
calculable and calculated nowadays.

{\it To summarize:} at least qualitatively, 
the phenomenology with a tachyonic gluon mass
which is quite large numerically stands well to a few highly nontrivial
tests. Further crucial tests of the model with the tachyonic 
gluon mass could be 
furnished with measurements of various correlation functions $\Pi_j(M^2)$
on the lattice.

\section*{Revisiting vacuum condensates: condensates 
of dimension $d=2$ in gauge theories.}

As is explained in
the first part of the lectures,
introduction of various vacuum condensates turned a useful way
to understand and characterize dynamics of QCD. 
The most famous 
example is the quark condensate:
\be
\langle 0|\bar{q}q|0\rangle \neq 0
\ee
where $q$ stands for light $u$- or $d$-quarks. 
In the realistic case of negligible small quark mass, a nonvanishing value
of the quark condensate manifests spontaneous breaking
of the chiral symmetry. 
 
Within the framework of QCD sum rules there emerged also the  
gluon condensate
$$
\langle 0|\alpha_s(G^a_{\mu\nu})^2|0\rangle \neq 0
$$
A non-vanishing value of this condensate does not signify breaking
of any symmetry but rather provides with a measure
of nonperturbative fields in the vacuum.
The gluon condensate appears to be the simplest gauge invariant quantity
characterizing nonperturbative vacuum fields. It has dimension $d=4$
and this dimension is echoed in the observation that
the leading non-perturbative corrections in the QCD
sum rules at large $Q$ are 
of order $\langle 0|\alpha_s(G^a_{\mu\nu})^2|0\rangle /Q^4$. 

Now, we are driven to a rather astonishing conclusion that
condensates of dimension $d=2$ may also have dynamical significance.
Indeed, the condensation of monopoles in 
QCD is described by $<\Phi_M^2>\neq 0$ which is a dimension $d=2$
vacuum expectation value.
Also, we saw that in the compact $U(1)$
there arises a nonvanishing string tension. 
The condensates are of nonperturbative nature and their appearance 
is related
to the compactness of the gauge group. Clearly, such
condensates cannot be detected by means of the standard OPE
since they are not related to softness of propagators in the Feynman graphs.

There is no direct clash with the gauge invariance since
$\Phi_M$ does not enter the Lagrangian but is constructed
in a rather indirect way, see the preceding section.
The situation when there emerge condensates 
not related at all to the original fields may appear, however, confusing.
In terms of the original fields condensates of dimension $d=2$
can be related only to 
the vacuum expectation value of $<A_{\mu}^2>$.
And in this section we will argue, following (Gubarev et. al., 1999)   
that the condensate $<A_{\mu}^2>$ constructed
directly on the vector-potential of the gauge field $A_{\mu}$ may also have
dynamical significance, upon some clarifications
Literally 
such a condensate is gauge non-invariant and devoid therefore of 
physical meaning. We will consider, however, the minimal value of
it, $<(A_{\mu}^2)_{min}>$.  Looking
for the minimal value of $A_{\mu}^2$
implies, of course, a particular choice of the gauge.
Also, we have to consider the Euclidean space-time so that the
minimal value of $A_{\mu}^2$ could be meaningful.

The relevance of $<(A_{\mu}^2)_{min}>$ can be understood on a simple
example of very thin cosmic strings borrowed from (Alford and Wilczek,
1989). Then the
field strength tensor cannot be detected directly. However, because of
of the Aharonov-Bohm effect particles can be scattered off
the strings. From our point of view, it is crucial that the system
is characterized by a non-vanishing $A_{\mu}$. Indeed, because
of the topological condition,
\be
\oint A_{\mu}dx^{\mu}~=~\int {\bf H\cdot }d{\bf s}~\equiv~\varphi
\ee
where $\varphi $ is the magnetic flux, the value of $A_{\mu}^2$ cannot
be brought down to zero everywhere. Thus, the value of $<(A_{\mu}^2)_{min}> \ne 0$ 
distinguishes
the system from the trivial vacuum for which, on the classical level,
one can choose $A_{\mu}=0$.

In case of the compact $U(1)$ gauge group, one can measure 
the $<(A_{\mu}^2)_{min}>$ numerically. Moreover, it is possible to substruct
explicitly the perturbative contribution to this quantity leaving only
the dynamically relevant part $<(A_{\mu}^2)_{min}>_{non-pert.}$.
The result for $<(A_{\mu}^2)_{min}>_{non-pert.}$ is presented
in Fig. 6. We see that the behavior of this condensate clearly
\begin{figure}[h]
  \begin{center}
    \leavevmode
    \epsfig{figure=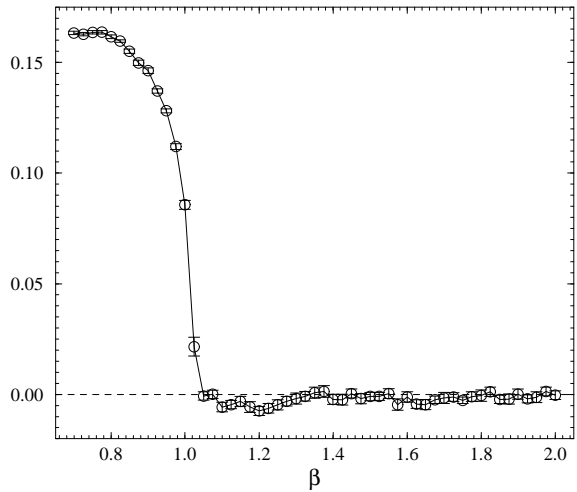,width=7.5cm}
    \caption{Non-perturbative contribution to $<(A_{\mu}^2)_{min}>$ as a function of
$\beta=1/e^2$. 
The plot is borrowed from (Gubarev et. al., 1999)}
    \label{fig:fig6}
  \end{center}
\end{figure}
signals the phase transition to the confining phase near
$e^2=1$. Although, it cannot be considered, strictly speaking, 
as an order parameter since even the nonperturbative part of 
$<(A^2_{\mu})_{min}>$  does not vanish in any phase.

{\it To summarize}: dimension two vacuum expectation values,
like $<(A_{\mu}^2)_{min}>$ may have a hidden topological meaning
and be dynamically significant.


\section{CONCLUSIONS.}

In Part I we considered power corrections within the OPE approach,
which culminates in the QCD sum rules (see, e.g., (\ref{qcd})).
The sum rules allowed to tune the power corrections 
to resonances in some channels,
like the $\rho$-meson channel. There are many further successful 
applications, see, e.g., Reinders et.al. (1985), Narison (1989),
which we just did not have time at all to cover. 
More recently the technique was generalized to observables
directly in the Minkowski space by using infrared renormalons or
infinitesimal gluon mass.

On the other hand, in some so to say exceptional channels
the sum rules certainly fail (Novikov et.al., 1981).
Most conspicuous cases are the $\pi$-meson, $0^+$-gluonium channels.
There are some explanations to the failures in terms of 
direct transitions of the currents to instantons (Novikov et.al. (1981),
Shuryak and Schafer (1998)). However, the instanton corrections are lacking
simple analytical structure. Also, neither the soft vacuum fields accounted
for by the OPE nor the direct instantons seem to encode the effects
of confinement.

Moreover, very recent studies of the power corrections in the
Abelian Higgs model reveal novel $1/Q^2$ corrections
which are associated with small distances and are not accounted for
in the operator product expansion.
If such corrections exist in the QCD case as well, then there
are quite a few places where the new corrections could be detected
through measurements, mostly through measurements on the lattice.
The existing data are not discouraging for the new corrections, not
at all. Further checks are desirable.

To summarize, the nature may turn to be generous as far as 
power corrections are concerned.
I am borrowing this term from a talk on dark matter. Indeed, first
people assumed that there should be a single dominant source
of the dark matter, and now it appears distributed among various
equally important components.
Similar picture may be true for the sum rules.
Indeed, very first idea would be that theoretically the 
correlation functions $\Pi_j(Q^2)$ could be found perturbatively at large $Q^2$
and
the growth of the effective coupling at smaller $Q^2$ would signal
the breaking of the asymptotic freedom. Then the picture got more involved
and the effect of soft non-perturbative fields was included
in terms of the quark and gluon condensates.
It appears not suffice to explain the peculiarities of all the channels
and the effect of direct instantons was invoked.
As the latest development, hypothetical $1/Q^2$ corrections
associated with short strings are established within
the Abelian Higgs model which is thought to mimic the QCD confinement.

Thus it appears now that practically all known ``ingredients'' of QCD
have already found their way
into the physics of the power corrections.

\section*{Acknowledgments}

We are acknowledging thankfully
discussions with M.N. Chernodub, A. Leonidov, V.A. Novikov, L. Stodolsky,
A.I. Vainshtein, M.I. Vysotsky. Work of F.V.G. and M.I.P. was partially supported by
grants RFBR 99-01230a and INTAS 96-370.

\newpage

\section*{References.}

Abrikosov, A.A., 
``On the magnetic properties of superconductors of the second group'',\\
{\it Sov.Physics JETP} {\bf 5}, 1174 (1957).\\

Aglietti, U. and Ligeti, Zoltan, ``Renormalons and confinement'' \\
{\it Phys.Lett.} {\bf B364}, 75(1995). \\

Akhoury, R. and  Zakharov, V.I., 
``Leading power corrections in QCD: from renormalons to phenomenology'',\\
{\it Nucl. Phys.} {\bf B465}, 295 (1996).\\

Akhoury, R. and  Zakharov, V.I.,
``A quick look at renormalons'',\\
{\it Nucl. Phys. Proc. Suppl.}
{\bf 54A} 217 (1997), (hep-ph/9610492).\\

Akhoury, R. and Zakharov, V.I., 
``On nonperturbative corrections to the
potential for heavy quarks'' \\
{\it Phys. Lett.} {\bf B438}, 165 (1998).\\

Alford, M.G. and Wilczek, Frank,
``Aharonov-Bohm interaction of cosmic strings with matter'', \\
{\it Phys. Rev. Lett.} {\bf 62}, 1071, (1989).\\

Badalian, A.M. and Morgunov, V.L.,
``Determination of $\alpha_s(1 \mbox{GeV})$ from the charmonium fine structure'',\\
E-Print Archive: hep-ph/9901430 (1999).\\

Bali, Gunnar S., Schilling, Klaus, and Wachter, Armin,\\
``Quark-anti-quark forces from $SU(2)$ and $SU(3)$
gauge theories on large lattices'', \\
E-print archive: hep-lat/9506017 (1995). \\

Bali, Gunnar S.,
``Do ultraviolet renormalons contribute to the QCD static potential?'',\\
E-mail archive, hep-ph/9905387 (1999).\\

Balitsky, Ya. Ya.,
``Wilson loop for the stretched contours in vacuum fields
and the small distance behaviour of the interquark potential'',\\
{\it Nucl. Phys.} {\bf B254}, 166 (1985).\\

Ball, James S. and Caticha, Ariel.
``Superconductivity: a testing ground for models of confinement'',\\
{\it Phys.Rev.} {\bf D37} 524, (1988). \\

Beneke, M., ``Renormalons'',\\
E-print archive: hep-ph/9807443 (1998).\\

Brandt, Fichard A., Neri, Filippe, and Zwanziger, Daniel,
``Lorentz invariance of the quantum field theory of 
electric and magnetic charge''\\
{\it Phys. Rev. Lett.} {\bf 40}, 147 (1978).\\

Buchmuller, W. and Tye, S.H.H.,
``Quarkonia and quantum chromodynamics'',\\
{\it Phys. Rev.} {\bf D24} 132 (1981).\\

Casimir, H.G.B. and Polder, D.,
``The influence of retardation on London-van der Waals Forces''\\
{\it Phys. Rev.} {\bf 73} 360 (1948). \\

Chen, D., Brower, R.C., Negele J.W., Shuryak, E.,
``Heavy quark potential in the instanton liquid model'',\\
e-Print Archive: hep-lat/9809091 (1998).\\

Chernodub, Maxim. N. and Polikarpov, Mikhail. I.,
''Abelian Projections and Monopoles''\\
In *Cambridge 1997, Confinement, duality, and nonperturbative aspects 
of QCD* 387-414. \\
(E-print archive: hep-lat/9710205).\\

Chernodub, Maxim. N., Gubarev, Fedor.V., Polikarpov, Mikhail. I.,
and Veselov, Alexander. I.,
``Monopoles in the Abelian projection of gluodynamics'',\\
{\it Prog. Theor. Phys. Suppl.} {\bf 131}, 309 (1998).
(E-print archive: hep-lat/9802036). \\
 
Chernodub M.N., Kato S., Nakamura N., Polikarpov M.I. and Suzuki T.,
"Various representations of infrared effective lattice $SU(2)$
gluodynamics",\\
preprint KANAZAWA-98-19, Feb 1999, 
E-Print Archive: hep-lat/9902013.\\ 

Chetyrkin, K.G. and Spiridonov, V.P., 
``Nonleading mass corrections and renormalization of the operators
$m\bar{\psi}\psi$ and $G_{\mu\nu}^2$.''\\
{\it Sov. J. Nucl. Phys.} {\bf 47}, 522 (1988).\\

Chetyrkin, K.G., Narison, Stephan, and Zakharov, V.I., 
``Short distance tachyonic mass and $1/Q^2$ corrections'',\\
e-Print Archive: hep-ph/9811275  (1998).\\

Chu M.C., Grandy J.M., Negele J.W,, and Huang S,. 
``Lattice calculation of point-to-point current correlation
functions in the QCD vacuum'',\\
{\it Phys. Rev. Lett.} {\bf 70}, 255 (1993). \\

David, F.,
``On the ambiguity of composite operators, I.R. renormalons
and the status of the operator product expansion''\\
{\it Nucl. Phys.} {\bf B234}, 237 (1984).\\

Del Debbio, L., Faber, M., Greensite, J. and Olejnik, S.\\
``Some cautionary remarks on abelian projection and abelian dominance'',\\ 
{\it Nucl.Phys.Proc.Suppl.} {\bf 53} (1997) 141;\\
E-print archive: hep-lat/9607053.\\

Di Giacomo, Adriano,
``Monopole condensation and quark confinement'',\\
{\it Prog. Theor. Phys. Suppl.} {\bf 131} (1998) 161. (E-print archive:
hep-lat/9803008).\\ 
 
Dokshitzer, Yu.L., Marchesini, G., Webber, B.R.,
``Dispersive approach to power behaved contributions in 
QCD hard processes'',\\
{\it Nucl. Phys.} {\bf B469}, 93 (1996).\\

Dosch, H.G. and Simonov, Yu.A.,
``The area law of the Wilson loop and vacuum field correlators'',\\
{\it Phys. Lett.} {\bf B205}, 339 (1988).\\

Faber, M., Greensite, J., and Olejnik, S.,
``Center projection with and without gauge fixing'',\\
{\it JHEP} {\bf 9901}, 008 (1999).\\

Fingberg, Jochen,
``Super-heavy quarkonia and gluon condensate'',\\
E-print archive: hep-lat/9810050.\\

Geshkenbein, B.V. and Ioffe, B.L. 
``The role of instantons in generation of mesonic mass spectrum'',\\
{\it Nucl. Phys.} {\bf B166} (1980) 340.\\

Gubarev, F.V., Polikarpov, M.I., and Zakharov, V.I.,
``Short strings in Abelian Higgs model'',\\
E-print archive, hep-th/9812030 (1998). \\

Gubarev, F.V., Stodolsky, L., and Zakharov, V.I.,
``On condensates dimension $d=2$ in gauge theories''\\
in preparation, (1999).\\

Huber, Stephan J., Reiter, Martin, and Schmidt Michael G.,\\
``A tachyonic gluon mass: between infrared and ultraviolet'',
E-print- archive: hep-ph/9906358 (1999).\\

Ivanenko T.L., Pochinsky A.V. and Polikarpov M.I., 
"Extended abelian monopoles and confinement in the $SU(2)$
lattice gaugetheory",\\
{\it Phys.Lett.} {\bf B252}, 631 (1990).\\ 

Jersak, J., Neuhaus, T., Pfeiffer H., 
``Scaling analysis of the magnetic monopole mass and condensate
in the pure $U(1)$ lattice gauge theory'', \\
e-Print Archive: hep-lat/9903034. (1999).\\ 

Kataev, A.L., Parente, G., and Sidorov, A.V.,
``Higher twists and $\alpha_s(M(Z))$ from the NNLO
QCD analysis of the CCFR data for the $xF_3(x)$ structure function'',\\
e-Print Archive: hep-ph/9905310 (1999).\\ 

Kato S., Nakamura N., Suzuki T., and Kitahara S.,
``Perfect monopole action for infrared $SU(2)$ QCD'',\\
{\it Nucl.Phys.} {\bf B520}, 323 (1998).\\

Leutwyler, H., 
``How to use heavy quarks to probe the QCD vacumm'',\\
{\it Phys. Lett.} {\bf B98}, 447 (1981). \\

Mandelstam, S., 
``Vortices and quark confinement in nonabelian gauge theories'',\\
{\it Phys. Lett.} {\bf 53B}, 476 (1975). \\

Mueller, A.H.,
``Renormalons and phenomenoly in QCD'',\\
{\it Nucl. Phys.} {\bf B250}, 327 (1985).\\

Nambu, Y,
``Strings, monopoles and gauge fields'',\\
{\it Phys. Rev.} {\bf D10}, 4662 (1974).\\

Narison, S., 
{\it QCD Spectral Sum Rules},\\
World Scientific, Singapore (1989).\\

Narison, Stephan,
``Determination of the $D=2$ 'operator' from $e^+e^-$ data''.\\
{\it Phys. Lett.} {\bf B300}, 293 (1993). \\

Nielsen, H.B. and Olesen, P.,
``Vortex-line models for dual strings'',\\
{\it Nucl. Phys.} {\bf B61}, 45 (1971).\\

Novikov, V.A., Okun, L.B., Shifman, M.A., Vainshtein, A.I., 
Voloshin, M.B., and Zakharov, V.I.,
``Charmonium and gluons'',\\
{\it Phys.Rept.} {\bf 41}. 1 (1978).\\

Novikov, V.A., Shifman, M.A., Vainshtein, A.I., and Zakharov, V.I.,
``Are all hadrons alike?'',\\
{\it Nucl. Phys.} {\bf B191}, 301 (1981).\\

Novikov, V.A., Shifman, M.A., Vainshtein, A.I., and Zakharov, V.I., 
``Wilson's operator expansion: can it fail?'',\\
{\it Nucl. Phys.} {\bf B249}, 445 (1985).\\

Peter, Markus, 
``The static potential in QCD: a full two loop calculation'',\\
{\it Nucl. Phys.} {\bf B501}, 471 (1997). \\

Polyakov, A.M.,
``Particle spectrum in the quantum field theory'',\\
{\it JETP Lett.} {\bf 20} 194, (1974).\\

Polyakov, A.M.,
``Compact gauge fields and infrared catasrophe'',\\
{\it Phys. Lett.} {\bf B59}, 82 (1975).\\

Reinders, L.J., Rubinstein, H., and Yazaki, S.,
``Hadron properties from QCD sum rules'',\\
{\it Phys. Rep.} {\bf 127,} 1 (1985).\\

Simonov, Yu. A., 
``Confinement'',\\
{\it Phys. Usp.} {\bf 39}, 313, (1996),\\

Simonov, Yu. A.,
``Perturbative-nonperturbative interference in the static
QCD interaction at small distances'',\\
E-print archive: hep-ph/9902233, (199).\\

Shifman, M.A., Vainshtein, A.I., Zakharov, V.I.,
``Photoproduction of charmed particles and 
asymptotically free field theories'',\\
{\it Phys. Lett.} {\bf B65}, (1976) 255.\\

Shifman, M.A., Vainshtein, A.I., and Zakharov, V.I.,
``QCD and resonance physics'',\\
{\it Nucl. Phys.} {\bf B147}, 385, 448, 519 (1979).\\

Shuryak, E. and Schafer, T., 
``The QCD vacuum as an instanton liquid'',\\
{\it Ann. Rev. Nucl. Part. Sci.}, {\bf 47} (1997) 359.\\

Voloshin, M.B., 
``On dynamics of heavy quarks in a nonperturbative QCD vacuum'',\\
{\it Nucl. Phys.} {\bf 154}, 365 (1979).\\

't Hooft, G.,
``Magnetic monopoles'', {\it Nucl. Phys.} {\bf B79} 276 (1974).\\

't Hooft, G.,
``Gauge theory for string interactions'',\\
{\it in} "High Energy Physics", Proc. EPS Intern. Conf.,
ed A. Zichichi, Editrici Compositori, (1976).\\

't Hooft, G.,``Topology of the gauge condition and new confinement phases
in nonabelian gauge theories'',\\
{\it Nucl. Phys.} {\bf B190}, 455 (1981).\\

Webber, B.R., 
``Renormalon phenomena in jets and hard processes'',\\
{\it Nucl. Phys. Proc. Suppl. }{\bf 71} 66 (1999) 
(e-Print Archive: hep-ph/9712236).\\ 

Wilson, K., ``On Products of Quantum Field Operators at Short Distances''\\
Cornell Report, (1964) and {\it Phys. Rev.} {\bf 179} 1499 (1969).\\
 
Yndurain, F.J.,``Bound states of heavy quarks in QCD'', \\
{\it Nucl. Phys. Proc. Suppl.} {\bf 64}, 433 (1998).\\

Zakharov, V.I.,``Renormalons as a bridge between perturbative and 
nonperturbative physics'',\\ { \it Prog. Theor. Phys. Suppl.} 
{\bf 131}, 107 (1998) (hep-ph/9802416).\\

Zwanziger, Daniel,``Local Lagrangian quantum field theory of electric 
and magnetic charges'',\\
{\it Phys. Rev.} {\bf D3} 880 (1971).

\end{document}